\documentclass[a4paper,11pt]{article}

\usepackage[T1]{fontenc}
\usepackage{jheppub}
\preprint{USTC-ICTS/PCFT-26-58}

\usepackage{mathtools}
\usepackage{booktabs}
\usepackage{array}
\usepackage{longtable}
\usepackage{microtype}
\usepackage{enumitem}

\allowdisplaybreaks[2]
\setlist[enumerate]{leftmargin=8mm,itemsep=2pt,topsep=3pt}

\newcommand{\appref}[1]{appendix~\ref{#1}}
\newcommand{\secref}[1]{section~\ref{#1}}
\newcommand{\figref}[1]{figure~\ref{#1}}

\newcommand{\e}{\mathrm e}
\newcommand{\dd}{\mathrm d}
\newcommand{\Li}{\operatorname{Li}}

\usepackage{tikz}
\usetikzlibrary{arrows.meta,decorations.pathmorphing}
\title{The large $N$  vector model  with angular velocity}
\author[a]{Justin R. David}
\author[b,c]{, Srijan Kumar}
\affiliation[a]{\vspace{.1cm} Centre for High Energy Physics,  Indian Institute of Science,
	C. V. Raman Avenue, Bangalore 560012, India. \vspace{.1cm} }
\affiliation[b]{ Interdisciplinary Center for Theoretical Study, University of Science and Technology of China, Hefei,
	Anhui 230026, China.\vspace{.1cm} }
\affiliation[c]{Peng Huanwu Center for Fundamental Theory, Hefei, Anhui 230026, China.}

\emailAdd{justin@iisc.ac.in}
\emailAdd{srijankumar@ustc.edu.cn}

\abstract{We study the free energy of a critical vector model at large $N$  on
	$S^{1}\times S^{2}$ with an angular velocity $\hat\mu$ without the singlet constraint. We study the model for which the  large $N$ dynamics is controlled by the  uniform saddle point of the auxiliary field arising in the Hubbard-Stratonovich transformation. The
	leading high-temperature behaviour is determined analytically both as an expansion about $\hat\mu r=0$ and $\hat\mu^{2}r^{2}=1$
	where $r$ is the radius of the sphere.
	We supplement the analytic results with a numerical analysis that agrees with both the analytical expansions in their respective
	regimes and smoothly interpolates between them. The leading
	high-temperature contribution to the free energy develops a pole
	at $\hat\mu^{2}r^{2}=1$, in agreement with expectations from the thermal effective
	field theory. Its residue coincides with that of the massless  free theory. 
	Sub-leading terms, however, exhibit non-analytic dependence on the angular
	velocity and distinguish the critical fixed-point result from the
	free theory answer.
	The residue at the pole can also be obtained  by placing the  model 
	on the pp-wave geometry. We show that the residue agrees  with that obtained from  the direct computation. The free energy of the model connects the non-trivial fixed point of the $O(N)$ model at $\hat\mu r=0$ to its free fixed point at $\hat\mu^2r^2=1$.}

\begin{document}
	\maketitle

	\section{Introduction}
	\label{sec:introduction}

	The
	study of  conformal field theories at finite temperature is important in many phenomenological contexts both in high energy physics and condensed matter physics.  From the theoretical point of view a  reason for their importance  is because  the physics of 
	conformal field theories at finite temperature  at strong coupling can be related to black hole physics 
	by  the AdS/CFT correspondence.
	However,   non-zero temperature breaks  some of the symmetries  of the conformal group 
	making it less constrained as compared to its zero temperature counterpart making its study more involved. 
	In spite of this there has been a  lot of progress  in using symmetries of the conformal group to constrain 
	and determine observables  in thermal CFT's \cite{Iliesiu:2018fao,Gobeil:2018fzy,Iliesiu:2018zlz,Petkou:2018ynm,David:2023uya,Marchetto:2023fcw,Karydas:2023ufs,Petkou:2026upo,Diatlyk:2023msc,Benedetti:2023pbt,David:2024naf,Kumar:2025txh,Marchetto:2023xap,Barrat:2025wbi,Barrat:2025nvu,Niarchos:2026vow}.
	Here the large $N$ vector models played an instrumental role providing a testing ground for verifying the methods 
	developed in the thermal bootstrap program.    With the  motivation of
	understanding CFT's at finite temperature,  new results in  this model have been found.  These involve  evaluation
	of partition functions as well as one point functions of vector models on 
	curved geometries \cite{David:2024pir,David:2025tqn,Mauro:2026zus,Parmentier:2026aqh}. Vector models provide non-trivial examples to benchmark the general CFT results in curved geometries at finite temperature \cite{Buric:2024kxo,Buric:2025uqt,Ammon:2025cdz}. 
	\\
	
	%
	The thermodynamics of rotating black holes in $AdS$ provides the motivation to study conformal field theories 
	in the presence of angular potential. 
	Early works led to constraining the partition functions of conformal field theories with finite angular velocities 
	using a thermal effective action \cite{Bhattacharyya:2007vs,Banerjee:2012iz,Jensen:2012jh,Shaghoulian:2015lcn}. 
	Study of the large $N$ vector models on $S^1\times S^2$ with non-zero angular potential is an interesting problem in this regard as a direct field theory computation for a non-trivial CFT. It can provide an important test for the predictions obtained from the thermal effective action. The non-zero angular potential breaks the $SO(3)$ symmetry to $O(2)$. Thus, unlike at zero angular potential, the large-$N$ dynamics of the conventional $O(N)$ model is not fully captured by the constant mode of the auxiliary-field in the Hubbard-Stratonovich transformed action. Hence the analysis is technically more involved.

	In this paper we will restrict our attention to a critical vector model at large $N$  on $S^1\times S^2$ where the large $N$ dynamics only depends on the constant mode of the auxiliary field even at non-zero angular potential. The partition function of the model is given as below involving the constant mode  on the $S^2$ of the auxiliary field $\zeta_0$
	\begin{align}
		Z(\beta, \hat \mu)
		={}&\int d\zeta_0\,\mathcal D\phi\,
		\exp\Bigl\{-\frac12\int d^3x\,\sqrt g\,\Bigl[
		g^{\mu\nu}(D_\mu\phi_i)(D_\nu\phi_i)
		+\Big(\frac{\mathcal R}{8}-i\zeta_0\Big)\phi_i\phi_i
		+\frac{N}{4\lambda}\zeta_0^2
		\Bigr]\Bigr\},
		\nonumber
		\\
{\rm with}\qquad		D_\mu&=\nabla_\mu+\hat \mu\,\delta_{\mu\tau}L_z,
		\qquad L_z=-i\partial_\varphi \label{eq:zero-mode-path-integral-intro} .
	\end{align}
	Here, if the auxiliary field  depended on all the modes on $ S^2$, the model would be the $O(N)$ model with quartic interaction. 
	$\phi_i$ is the real scalar transforming as the fundamental of $O(N)$. 
	$L_z$ is the angular momentum along the $z$ axis with an angular potential $\hat \mu$, 
	 $\beta$ the inverse temperature.  ${\cal R}$ is the Ricci  scalar of $S^2$   with radius $r$.
		Therefore the  model we consider is a simplified
	 mean field version of the original model, but as we will see the behaviour of the partition function 
	of the model considered in this paper is interesting. 
	
	Recently, there has been a renewed effort at characterising  the behaviour of conformal field theories at finite temperature and 
	angular velocity  \cite{Shaghoulian:2016gol,Benjamin:2023qsc,Allameh:2024qqp,Anand:2025mfh}.
	These papers, in particular the work of  \cite{Anand:2025mfh} have led to the conjecture 
	that  thermal free energies of all conformal field theories develop a pole at  $\hat \mu^2r^2=1$, where $r$ is the radius of the sphere. 
	For conformal field theories   in $d=2$ space-time dimensions, the existence of this pole can be shown using modular invariance \cite{Shaghoulian:2015lcn,Benjamin:2023qsc,Anand:2025mfh}.  The residue at the pole has been shown to be proportional to the central charge of the theory. 
	For conformal field theories on $S^1\times S^{d-1}$ with $d\geq 3$  
	the existence of the pole  has been argued from thermal effective field theory \cite{Anand:2025mfh}  and is viewed
	as a generalization to the Cardy formula in higher dimension 
	\footnote{Earlier work on effective  field theory  techniques applied in supersymmetric theories  also predicted the existence of the 
		pole \cite{DiPietro:2014bca,Cassani:2021fyv}.}.
	The  residue 
	at this  pole depends on the theory under consideration. 
	In \cite{Komargodski:2026ain}, it was argued that this residue can be extracted from evaluating  the partition function of the 
	conformal field theory on the pp-wave geometry.  In  \cite{Lee:2026azy}  the existence of the pole was shown to hold in a class 
	non-relativistic conformal theories.   Free energies of 
	conformal higher derivative and higher spin fields also show poles as the angular velocity approaches unity  \cite{Mukherjee:2026dfu}. 
	
	In this work we will provide further evidence for the conjecture that partition functions of conformal field theories develop
	poles as the angular velocity approaches unity, in units of the radius of the sphere. 
	We evaluate the free energy of the large $N$ critical vector model described in \eqref{eq:zero-mode-path-integral-intro} at high temperature  at arbitrary $\hat \mu r$  in $d =3$ and
	show that 
	as $\hat\mu^2 r^2  \rightarrow 1$, the free energy develops a pole.
	This theory is  in the non-singlet sector and is a  non-trivial interacting thermal  conformal field theory. An important result of this study is as follows.   We show that the free energy of the model in (\ref{eq:zero-mode-path-integral-intro})  smoothly 
	interpolates between the non-trivial fixed point of the standard 
	 $O(N)$ model  at large $N$  with quartic interaction  and vanishing angular potential, 
	 $\hat\mu r=0$ to its free fixed point at $\hat \mu^2r^2=1$.

	
	\subsubsection*{Summary of the results}
	
	We summarize the  main observations of this paper  of the vector model \eqref{eq:zero-mode-path-integral-intro} in presence of the angular potential in $(2+1)\, d$.  
	The model admits two fixed points one is the  free fixed point  and the other  is a  strong coupling  fixed point.
	At the  free fixed point the high temperature contribution to the free energy is known to have the following form \footnote{Throughout the paper we factor out the $N$ that occurs as a pre-factor in the free energy. }
	\begin{equation}
		\frac{\beta^2}{r^2}\log Z_{\rm free}^{(0)}(\hat{\mu} r)
		=\frac{2\zeta(3)}{1-\hat \mu^2r^2}.
		\label{eq:free-leading-pole-intro}
	\end{equation}
	The superscript ``(0)'' denotes the leading high temperature contribution. Thus the leading high temperature behaviour of the free energy is given by the pole at $\hat\mu^2 r^2=1$.  Let us write this free energy as an expansion about $\hat{\mu} r=0$, we obtain
	\begin{align} \label{expfreeir}
	\frac{\beta^2}{r^2}	\log Z^{(0)}_{\rm free}(\hat\mu r)={2\zeta(3)}\Big(1+\hat\mu^2r^2+\hat{\mu}^4r^4+O\big((\hat\mu r)^6\big) \Big). 
	\end{align}
	The non-trivial  fixed  point is characterised by a  thermal mass which arises due to the uniform saddle point of the auxiliary field in the Hubbard-Stratonovich transform of the quartic action at  the strict large $N$ limit.
	The thermal mass   $\tilde m(\hat\mu r)$,  which satisfies  the gap equation at small $\hat \mu r$  takes the following form
	\begin{align}
		&\frac{\tilde m(\hat\mu r)\beta}{2\log\rho_{\rm g}}
		=1+\frac{\hat\mu^2r^2}{3}
		+\frac{45+2\sqrt5\log\rho_{\rm g}}{225}\hat\mu^4r^4
		+O\big((\hat\mu r)^6 \big)
		\label{eq:mass-x1-series-intro},\quad {\rm with }\ \rho_{\rm g}\equiv(1+\sqrt5)/2,  \nonumber \\ 
		& \qquad\qquad \qquad {\rm with} \qquad  \hat \mu r \ll1, 
	\end{align}
	The free energy as  an expansion  in small $\hat \mu r  $  is given by 
	\begin{align}
		&\frac{\beta^2}{r^2}\log Z_{\rm saddle}^{(0)}(\hat\mu r)
		=\frac85\zeta(3)+\frac85\zeta(3)\hat\mu^2 r^2
		+\Big(\frac85\zeta(3)
		+\frac{4\sqrt5\log^3\rho_{\rm g}}{45}\Big)\hat\mu^4r^4
		+O((\hat\mu r)^6).
		\label{eq:F-x1-final-intro}
	\end{align}
	Observe the difference in the expansion of the free energy 
	at the non-trivial fixed point vs the free fixed point from (\ref{expfreeir}) and (\ref{eq:F-x1-final-intro}).

	Let us now present the results for 
	the  thermal mass $\tilde m(\hat\mu r)$  of the vector model at large $N$ which satisfies the gap equation at $\hat \mu^2 r^2=1$. 
	The expansion   takes the following form
	\begin{align}\label{mb mbx-intro}
		\tilde m(\hat \mu r)\beta\sim{}&\log^2\!\frac1{\sqrt{1-\hat\mu^2r^2}} . 
	\end{align}
	The nature of this expansion is non-analytic including the higher order terms which are presented in (\ref{eq:mass-corrected}). 
	This  solution for thermal mass, leads to the following expansion for the free energy about $\hat \mu^2 r^2 =1$.
	\begin{align} \label{intro-exp-irpartition}
		\frac{\beta^2}{r^2}\log Z_{\rm saddle}^{(0)}(\hat \mu r)
		\sim{}&\frac{2\zeta(3)}{1-\hat\mu^2r^2}
		-\frac16\log^6\!\frac1{\sqrt{1-\hat\mu^2r^2}}. 
	\end{align}
	The leading behavior at $\hat \mu^2r^2=1$ is given by a simple pole,   the residue at the pole coincides with the free theory answer \eqref{eq:free-leading-pole-intro}. The existence of the pole  confirms the expectation from thermal effective field theory,  the expansion 
	away from $\hat \mu^2r^2=1$ is non-analytic and there are sub-leading logarithmic divergent terms. 
	The  ratio of the free energy at the non-trivial fixed point to that at the free fixed point in  these two limits are given by 
	\begin{align}
		\lim_{\hat\mu r\to0} \frac{\log Z^{(0)}_{\rm saddle}}{\log Z^{(0)}_{\rm free}}=\frac{4}{5},\qquad {\rm and }\qquad  	\lim_{\hat\mu^2 r^2\to1} \frac{\log Z^{(0)}_{\rm saddle}}{\log Z^{(0)}_{\rm free}}=1.
	\end{align}
	\begin{figure}[tbp]
		\centering
		\includegraphics[width=0.88\textwidth]{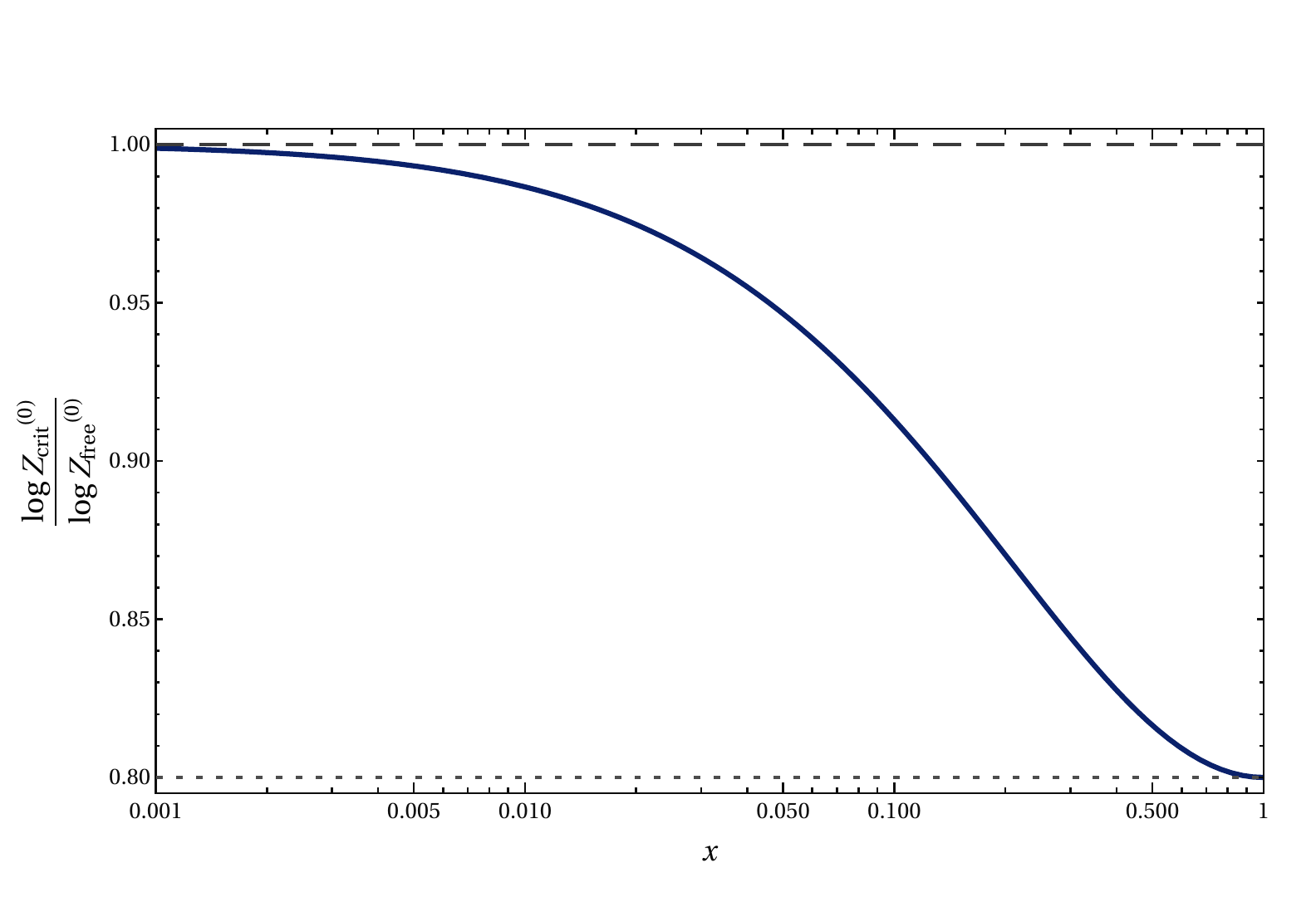}
		\caption{Ratio of the leading high-temperature free energy \eqref{eq:F-fixed} of the
			critical large-$N$ theory evaluated at the value of the thermal mass satisfying \eqref{eq:gap-fixed}, to that of the
			free-theory \eqref{eq:free-leading-pole}, shown with a logarithmic
			horizontal axis.  The curve is evaluated numerically, with $x^2=1-\hat\mu^2r^2$. It approaches unity as $x\to0$ and equals $4/5$ at $x=1$; the dashed and
			dotted horizontal lines mark these two limiting values.}
		\label{fig:interacting-free-ratio}
	\end{figure}
	In   figure \ref{fig:interacting-free-ratio},   this  ratio is obtained numerically  for  all values of $\hat \mu^2  r^2 $  between $0$ and unity. 
	We see that our model provides a
	smooth interpolation  between the non-trivial fixed point of the usual $O(N)$ model at $\hat\mu r=0$ and its free fixed point at $\hat\mu^2r^2=1$.
	%
	%
	
	As we mentioned earlier, in \cite{Komargodski:2026ain} it was argued 
	that the residue of the pole when $\hat \mu^2 r^2  \rightarrow 1$ can be obtained by
	considering the conformal field theory in the pp-wave geometry. 
	We consider the  vector model \eqref{eq:zero-mode-path-integral-intro}  at large $N$ in the pp-wave geometry and show that the only   critical point it admits is 
	when the theory is free. Therefore the residue of the pole must coincide with that of the free theory. 
	This provides a consistency check of our results in  (\ref{eq:free-leading-pole-intro}) and (\ref{intro-exp-irpartition})
	from which we see residues  of the pole  of the free theory coincides 
	with that of the critical vector model.

	The rest of the paper is organized as follows. In \secref{sec:free} we review the free scalar with the angular potential. Section \ref{sec:interacting} has the calculations for free energy of this vector model at large $N$, both at $\hat \mu r=0$ as well as at $\hat \mu^2r^2=1$. Section \ref{sec:plane-wave} has the calculation of the free energy of the model on the plane wave geometry. Section \ref{sec:conclusions} has our discussions. Appendix \ref{app:free-highT} computes the high temperature asymptotics of the free theory pole contribution. Appendix \ref{app:bessel-poisson} computes the small $\tilde m\beta x$ expansion of the gap equation. In the appendix \ref{app:thermal-mass-orders} we evaluate the thermal mass as an expansion about $\hat \mu^2r^2=1$ till a few higher orders.
Finally appendix \ref{Non-uniform} shows the non-uniformity of the saddle point in the auxiliary field for the conventional $O(N)$ model.

\section{Free theory}
\label{sec:free}

We begin with a single real conformal scalar field on
$S^1_\beta\times S^2_r$. 
The action involving a free  scalar field coupled with the curvature of the background reads as follows
\begin{equation}
 S_{\rm free}[\phi]
 =\frac12\int_0^\beta\dd\tau\int_{S^2}\dd^2x\,\sqrt g\,
 \left[
 (\partial_\tau\phi)^2+g^{ab}\nabla_a\phi\nabla_b\phi
 +\frac{\mathcal R}{8}\phi^2
 \right],
 \qquad \mathcal R=\frac{2}{r^2}.
 \label{eq:free-action}
\end{equation}
Rotation is introduced by coupling the thermal ensemble to the
angular momentum $L_z$, taken along the $z$-axis for convenience.  For an imaginary angular velocity $\hat\mu=i\mu$,\footnote{We adopted the convention of imaginary angular potential following the convention of imaginary $U(1)$ chemical 
	potential in the Euclidean theory. However all expressions in $\mu$  can be converted to  the real angular potential 
	by the simple replacement of $\mu = - i \hat \mu$. } the
corresponding grand-canonical partition function is
\begin{equation}
 Z_{\rm free}(\beta,\mu)
 =\operatorname{Tr}\!\left[\e^{-\beta(H-i\mu L_z)}\right].
 \label{eq:free-trace}
\end{equation}
To make the effect of this angular potential on the Euclidean action
explicit, we first write the trace in canonical form.  Rotational invariance
implies $[H,L_z]=0$, while $L_z=-i\partial_\varphi$ on one-particle
wavefunctions and
$L_z=\int_{S^2}\dd^2x\,\sqrt g\,\pi\,\partial_\varphi\phi$ in canonical
variables.  The trace \eqref{eq:free-trace} therefore admits the exact
Euclidean path integral representation
\begin{align}
 Z_{\rm free}(\beta,\mu)
 &={}\int\mathcal D\phi\,\mathcal D\pi\,
 \nonumber\\[-1mm]
 &\quad\times
 \exp\Big[-\int\dd^3x\,\sqrt g
 \Big(
 \frac12\pi^2
 -i\pi\bigl(\partial_\tau+\mu\partial_\varphi\bigr)\phi
 +\frac12 g^{ab}\nabla_a\phi\nabla_b\phi
 +\frac{\mathcal R}{16}\phi^2
 \Big)\Big],\nonumber\\
&\qquad\qquad {\rm with} \qquad\phi(\tau+\beta)=\phi(\tau).
 \label{eq:free-phase-space}
\end{align}
The periodic boundary condition displayed in
\eqref{eq:free-phase-space} implements the bosonic thermal trace.  Completing
the square in $\pi$ in \eqref{eq:free-phase-space} and performing its Gaussian
integral converts the angular coupling into a shift of the derivative along
the thermal circle.
The resulting configuration-space action is
\begin{equation}
\begin{aligned}
 S_{\rm free}[\phi;\mu]
 & =\frac12\int_0^\beta\dd\tau\int_{S^2}\dd^2x\,\sqrt g\,
 \left[
 g^{\mu\nu}(D_\mu\phi)(D_\nu\phi)
 +\frac{\mathcal R}{8}\phi^2
 \right],
 \\
 D_\mu
 & \equiv\nabla_\mu+i\mu\,\delta_{\mu\tau}L_z,
 \qquad L_z=-i\partial_\varphi .
 \label{eq:free-mu-action}
\end{aligned}
\end{equation}
Equation \eqref{eq:free-mu-action} gives
$D_\tau=\partial_\tau+\mu\partial_\varphi$, whereas the spatial
derivatives remain $D_a=\nabla_a$.  The angular potential can therefore be
viewed as a constant background connection supported only along
$S^1_\beta$.
The quadratic operator in \eqref{eq:free-mu-action} is diagonal in Fourier
modes along $S^1_\beta$ and spherical harmonics on $S^2_r$.  Thus we express the field as the following mode expansion
\begin{equation}
 \phi(\tau,\theta,\varphi)
 =\sum_{n\in\mathbb Z}\sum_{l=0}^{\infty}\sum_{m=-l}^{l}
 \phi_{nlm}\,\e^{i\omega_n\tau}Y_{lm}(\theta,\varphi),
 \qquad
 \omega_n=\frac{2\pi n}{\beta},
 \qquad
 L_zY_{lm}=mY_{lm}.
 \label{eq:free-mode-expansion}
\end{equation}
The angular potential shifts
the Matsubara frequency to $\omega_n+\mu m$.  Moreover, the eigenvalue
$l(l+1)/r^2$ of the scalar
Laplacian combines with the curvature coupling to give
$(l+\tfrac12)^2/r^2$.  Substitution of
\eqref{eq:free-mode-expansion} into \eqref{eq:free-mu-action}, followed by the
Gaussian integral over the mode coefficients, leads to the following expression
\begin{equation}
 \log Z_{\rm free}
 =-\frac12\sum_{n\in\mathbb Z}\sum_{l=0}^{\infty}\sum_{m=-l}^{l}
 \log\!\left[
 \left(\omega_n+\mu m\right)^2
 +\frac{\left(l+\tfrac12\right)^2}{r^2}
 \right].
 \label{eq:free-determinant}
\end{equation}
The  Matsubara sum in \eqref{eq:free-determinant} can now be
evaluated using the standard formula for the regulated sum found in 
\cite{Klebanov:2011uf}.  It recasts the free energy as a sum of a zero-point
contribution and a thermal logarithm, as shown below
\begin{equation}
 \log Z_{\rm free}
 =-\frac{\beta}{2r}\sum_{l=0}^{\infty}\frac{(2l+1)^2}{2}
 -\sum_{l=0}^{\infty}\sum_{m=-l}^{l}
 \log\!\left(
 1-\e^{-\frac{\beta}{r}(l+\frac12)+i\beta\mu m}
 \right).
 \label{eq:free-after-matsubara}
\end{equation}
The first term in \eqref{eq:free-after-matsubara} is the vacuum contribution.
For the conformal spectrum on $S^2$, its zeta-regularized value vanishes
because the sum over odd squares is proportional to $\zeta(-2)=0$
\begin{equation}
 -\frac{\beta}{2r}\sum_{l=0}^{\infty}\frac{(2l+1)^2}{2}
 =0.
 \label{eq:free-vacuum-zero}
\end{equation}
Thus the first term from the equation  \eqref{eq:free-after-matsubara} vanishes and only the 2nd term survives.
  We now expand the summand in the second term as a small $x$ expansion of the $\log (1-x)$ and the finite sum over $m$ is performed as a geometric sum.
\begin{align}
 \log Z_{\rm free}
 &=\sum_{n=1}^{\infty}\frac1n
 \sum_{l=0}^{\infty}\e^{-n\beta(l+\frac12)/r}
 \sum_{m=-l}^{l}\e^{in\beta\mu m}
 \nonumber,\\
 &=\sum_{n=1}^{\infty}\frac1n
 \sum_{l=0}^{\infty}\e^{-n\beta(l+\frac12)/r}
 \frac{\sin[(l+\frac12)n\beta\mu]}{\sin(n\beta\mu/2)}.
 \label{eq:free-angular-sum}
\end{align}
The remaining sum over the angular momentum $l$ in
\eqref{eq:free-angular-sum} is again a geometric sum.  Its evaluation leads to the
following compact exact form.  
\begin{equation}
 \log Z_{\rm free}
 =\sum_{n=1}^{\infty}
 \frac{\cosh\!\left(\frac{n\beta}{2r}\right)}
 {n\left[\cosh\!\left(\frac{n\beta}{r}\right)-\cos(n\beta\mu)\right]}.
 \label{eq:free-exact-real}
\end{equation}
 Now we use the following definition
\begin{equation}
 x^2\equiv 1+\mu^2r^2,
 \qquad
 0<x<1.
 \label{eq:free-x-def}
\end{equation}
Under the parametrization \eqref{eq:free-x-def}, the endpoint
$\mu^2r^2=-1$ corresponds to $x\to0$.  At fixed $\beta/r$, the 
summand in \eqref{eq:free-exact-real} admits a double pole at $x=0$, as demonstrated in the following expansion
\begin{equation}
 \log Z_{\rm free}
 =\frac{1}{x^2}\sum_{n=1}^{\infty}
 \frac{r}{\beta n^2\sinh\frac{\beta n}{2r}}
 \left\{
  1
  +\frac{x^2}4\left[
   \frac{\beta n}{r}\coth\!\left(\frac{\beta n}{r}\right)-1
  \right] +O(x^4)
 \right\}
.
 \label{eq:free-small-x-full}
\end{equation}
The $n$-sums in the above equation are convergent, but do not admit known closed form expression. The sum in the first term gives  the residue at the pole at $x^2=0$ and it reorganizes itself as a small $\beta/r$ expansion as given below
\begin{align}
 \log Z_{\rm free}
 \sim\frac{1}{x^2}\Bigg[{}&\frac{2r^2}{\beta^2}\zeta(3)
 +\frac{1}{12}\log\!\left(\frac{\beta}{r}\right)
 +\zeta'(-1)+\frac{\log2-1}{12}
 \nonumber\\
 &-\frac{7}{34560}\frac{\beta^2}{r^2}
 -\frac{31}{58060800}\frac{\beta^4}{r^4}
 -\frac{127}{19508428800}\frac{\beta^6}{r^6}
 +\cdots\Bigg]
 +O(x^0).
 \label{eq:free-summed-highT-few}
\end{align}
Note the asymptotic nature of the above expansion and its  derivation from \eqref{eq:free-small-x-full} is shown in \appref{app:free-highT}.
Now keeping only the leading contribution at the high temperature in the residue of the pole at $x^2=0$, we have
\begin{equation}
 \log Z_{\rm free}^{(0)}
 =\frac{2r^2}{\beta^2}\frac{\zeta(3)}{x^2}.
 \label{eq:free-leading-pole}
\end{equation}
The superscript $(0)$ used in the above equation denotes the leading contribution at small $\beta/r $. This is the exact leading contribution at high temperature to the free energy in presence of angular potential characterized by $x$, which can be verified by expanding \eqref{eq:free-exact-real} directly at small $\beta/r$.

\section{Critical large $N$ vector model}
\label{sec:interacting}

We begin with the standard $O(N)$ vector model with quartic interaction on
$S^1_\beta\times S^2_r$, without imposing a singlet constraint.  The field
content consists of $N$ real scalars $\phi_i$, with $i=1,\ldots,N$,
transforming in the vector representation of $O(N)$; the  coupling strength of the quartic coupling is denoted by $\lambda$.
\begin{equation}
 S[\phi]
 =\int_0^\beta\dd\tau\int_{S^2}\dd^2x\,\sqrt g\,
 \left[
 \frac12\nabla_\mu\phi_i\nabla^\mu\phi_i
 +\frac{\mathcal R}{16}\phi_i\phi_i
 +\frac{\lambda}{2N}(\phi_i\phi_i)^2
 \right],
 \qquad \mathcal R=\frac{2}{r^2}.
 \label{eq:ON-action}
\end{equation}
The model undergoes simplifications at a large $N$ limit and  admits a nontrivial fixed point at
$\lambda\to\infty$, while $\lambda=0$ corresponds to the free CFT described
in \secref{sec:free}.
The angular potential $i\mu$ couples to the generator $L_z$ of rotations about the
polar axis.  The thermal partition function is
\begin{equation}
 Z(\beta,\mu)
 =\operatorname{Tr}\!\left[
 \exp\!\left\{-\beta\bigl(H-i\mu L_z\bigr)\right\}
 \right].
 \label{eq:ON-trace}
\end{equation}
Since the Hamiltonian defined by \eqref{eq:ON-action} is rotationally
invariant, $[H,L_z]=0$. Now this partition function can be evaluated as the following path integral as was done in the  \secref{sec:free}.
\begin{equation}
 Z(\beta,\mu)
 =\int\mathcal D\phi\,\exp\!\left[-S[\phi;\mu]\right],
 \label{eq:ON-path-integral}\qquad {\rm with } \ \ \phi(\tau+\beta)=\phi(\tau).
\end{equation}
The action appearing in \eqref{eq:ON-path-integral} is
\begin{equation}
 \begin{aligned}
 S[\phi;\mu]
 &=\int_0^\beta\dd\tau\int_{S^2}\dd^2x\,\sqrt g\,
 \left[
 \frac12g^{\mu\nu}(D_\mu\phi_i)(D_\nu\phi_i)
 +\frac{\mathcal R}{16}\phi_i\phi_i
 +\frac{\lambda}{2N}(\phi_i\phi_i)^2
 \right],
 \\
 D_\mu&=\nabla_\mu+i\mu\,\delta_{\mu\tau}L_z,
 \qquad L_z=-i\partial_\varphi .
 \end{aligned}
 \label{eq:ON-mu-action}
\end{equation}
\\

Now we apply the Hubbard-Stratonovich trick to recast the quartic interaction in \eqref{eq:ON-action} into a form which is quadratic in $\phi$ by introducing an auxiliary field $\zeta$ in the following manner.
\begin{align}
 Z(\beta,\mu)
 ={}&\int\mathcal D\zeta\,\mathcal D\phi\,
 \exp\Bigl\{-\frac12\int_0^\beta\dd\tau
 \int_{S^2}\dd^2x\,\sqrt g\,\Bigl[
 g^{\mu\nu}(D_\mu\phi_i)(D_\nu\phi_i)
 \nonumber\\
 &\hspace{35mm}
 +\Big(\frac{\mathcal R}{8}-i\zeta\Big)\phi_i\phi_i
 +\frac{N}{4\lambda}\zeta^2
 \Bigr]\Bigr\}.
 \label{eq:HS-path-integral}
\end{align}
Note that we do not keep track of the field-independent normalization arising due to the Gaussian path integral in the auxiliary field, as it is not important for the leading large $N$ calculations.
We decompose the auxiliary field into its zero mode and its nonzero-mode
part,
\begin{align}
    \zeta=\zeta_0+\tilde \zeta.
    \label{eq:zeta-mode-decomposition}
\end{align}
In absence of the angular potential an important simplification follows from the fact that only the zero mode
$\zeta_0$ contributes at the leading order in the large-$N$ limit, whereas the
nonzero modes $\tilde \zeta$ enter only at subleading orders. In contrast the non-zero angular potential obstructs such simplification as it breaks the $SO(3)$ symmetry of the sphere to $SO(2)$. Thus the leading large $N$ contribution to the partition function is not completely determined by the zero mode as we show in the appendix \ref{Non-uniform}.\\

To simplify the  model described above we will turn our attention to a model where we ignore the  non-zero modes
 to the leading large $N$ contribution to the partition function.  It will turn out that the partition function of this  simplified model exhibits interesting  behaviour as we vary  the angular potential. 
Thus the model we consider is described by the partition function \footnote{Integrating out the $\zeta_0$ results of the simplified model 
results in the quartic interaction $ \int d^3 x d^3 x'  \phi_i(x) \phi_i(x) \phi_j(x') \phi_j(x')  $, the model is therefore non-local. }
\begin{align}
	Z(\beta,\mu)
	={}&\int d\zeta_0\,\mathcal D\phi\,
	\exp\Bigl\{-\frac12\int_0^\beta\dd\tau
	\int_{S^2}\dd^2x\,\sqrt g\,\Bigl[
	g^{\mu\nu}(D_\mu\phi_i)(D_\nu\phi_i)
	\nonumber\\
	&\hspace{35mm}
	+\Big(\frac{\mathcal R}{8}-i\zeta_0\Big)\phi_i\phi_i
	+\frac{N}{4\lambda}\zeta_0^2
	\Bigr]\Bigr\}.
	\label{eq:zero-mode-path-integral}
\end{align}
Now the path integral over $N$ scalar fields $\phi_i$ can be performed exactly as path integrals of free massive scalars with $-i\zeta_0$ identified as the  scalar mass squared.
\begin{equation}
 Z(\beta,\mu)
 =\int\dd\zeta_0\,
 \exp\!\left\{N\left[
  \log Z(\zeta_0;\beta,\mu)
  -\frac{\beta\pi r^2}{2\lambda}\zeta_0^2
 \right]\right\}.
 \label{eq:zero-mode-integral}
\end{equation}
Here $ Z(\zeta_0;\beta,\mu)$ denotes the partition function  of a single 
scalar of mass $-i\zeta_0$. We use the following notation for this mass, called the thermal mass
\begin{equation}
 \zeta_{0}=i\tilde m^2,
 \label{eq:thermal-mass-definition}
\end{equation}
We evaluate the integral in \eqref{eq:zero-mode-integral} by using the method of steepest
descent, therefore gives the following expression, with $\tilde m$ understood
to take its value at the saddle point,
\begin{equation}
 \frac1N\log Z(\beta,\mu)
 =\log Z(\tilde m;\beta,\mu)
 +\frac{\beta\pi r^2}{2\lambda}\tilde m^4
 .
 \label{eq:ON-leading-effective-action}
\end{equation}
Where the following saddle point condition determines the thermal mass $\tilde m$ in the following manner
\begin{equation}
 \frac{\partial}{\partial \tilde m}
 \left[
 \log Z(\tilde m;\beta,\mu)
 +\frac{\beta\pi r^2}{2\lambda}\tilde m^4
 \right]
 =0.
 \label{eq:ON-finite-lambda-saddle}
\end{equation}
At the critical point, $\lambda\to\infty$,
\eqref{eq:ON-finite-lambda-saddle} reduces to
\begin{equation}
 \frac{\partial}{\partial \tilde m}\log Z(\tilde m;\beta,\mu)
 =0.
 \label{eq:critical-saddle-general}
\end{equation}

Using \eqref{eq:thermal-mass-definition} in \eqref{eq:HS-path-integral}, the
one-component quadratic action at fixed homogeneous mass is
\begin{equation}
 S_{\tilde m}[\phi]
 =\frac12\int_0^\beta\dd\tau\int_{S^2}\dd^2x\,\sqrt g\,
 \left[
 g^{\mu\nu}(D_\mu\phi)(D_\nu\phi)
 +\left(\frac{\mathcal R}{8}+\tilde m^2\right)\phi^2
 \right].
 \label{eq:massive-action}
\end{equation}
The single scalar partition function used in \eqref{eq:zero-mode-integral}, can be written as the following sum in Matsubara modes as well as angular modes.
\begin{equation}
 \log Z(\tilde m;\beta,\mu)
 =-\frac12\sum_{n\in\mathbb Z}\sum_{l=0}^{\infty}\sum_{m=-l}^{l}
 \log\!\left[
 \left(\frac{2\pi n}{\beta}+\mu m\right)^2
 +\frac{(l+\tfrac12)^2}{r^2}+\tilde m^2
 \right].
 \label{eq:interacting-determinant}
\end{equation}
Applying the same regularized Matsubara identity as in \secref{sec:free} to
\eqref{eq:interacting-determinant} yields
\begin{align}
 \log Z(\tilde m;\beta,\mu)
 =-\frac{\beta}{2}
 \sum_{l=0}^{\infty}\sum_{m=-l}^{l}
 \sqrt{\frac{(l+\frac12)^2}{r^2}+\tilde m^2}
-\sum_{l=0}^{\infty}\sum_{m=-l}^{l}
 \log\!\Big(
 1-\e^{-\beta\sqrt{\frac{(l+\frac12)^2}{r^2}+\tilde m^2}+i\beta\mu m}
 \Big).
 \label{eq:interacting-after-matsubara}
\end{align}
Rewriting \eqref{eq:interacting-after-matsubara}, we separate the vacuum and
thermal contributions according to
\begin{align}
 \log Z(\tilde m;\beta,\mu)
 ={}&-\frac12\log Z_1(\tilde m)
 +\frac12\log Z_2(\tilde m;\beta,\mu),
 \label{eq:interacting-split}
\end{align}
where
\begin{align}
 \log Z_1(\tilde m)
 &=\beta\sum_{l=0}^{\infty}\sum_{m=-l}^{l}
 \sqrt{\frac{(l+\tfrac12)^2}{r^2}+\tilde m^2},
\nonumber\\
{\rm and}\quad \log Z_2(\tilde m;\beta,\mu)
 &=2\sum_{l=0}^{\infty}\sum_{m=-l}^{l}
 \sum_{n=1}^{\infty}\frac1n\,
 \e^{-n\beta\sqrt{(l+\tfrac12)^2/r^2+\tilde m^2}+i\mu mn\beta}.
 \label{eq:Z2-def}
\end{align}
These sums do not admit compact closed form expressions but we provide the computation of these two terms given  in the above equation as high temperature expansions in the next subsection. 

\subsection{High temperature set-up}
\label{subsec:high-temperature-expansion}
Let us  first consider the term $\log Z_1(\tilde{m}) $ from \eqref{eq:Z2-def} which does not have an explicit dependence on $\mu$. This term admits  the following  systematic expansion at small
$\beta/r$, as was computed in  \cite{David:2024pir}.
\begin{align}
 &\log Z_1
 =\sum_{p=0}^{\infty}\log Z_1^{(p)},
 \nonumber\\
{\rm where}\qquad &\log Z_1^{(p)}
 \equiv
 \frac{(-1)^p(4p-2)}{4\pi(2p)!}\,
 B_{2p}(\tilde m\beta)^{3-2p}
 \left(\frac{\beta}{r}\right)^{2p-2}
 \Gamma\!\left(p-\frac32\right)
 \Gamma\!\left(p+\frac12\right).
 \label{eq:Z1-highT}
\end{align}
In \eqref{eq:Z1-highT}, $B_{2p}$ denotes the Bernoulli number and
$\log Z_1^{(p)}$ is of order $(\beta/r)^{2p-2}$. In all the further discussions we will only focus on the leading term from the above expansion in small $\beta/r$, thus we write the leading term explicitly
\begin{equation}
 \log Z_1^{(0)}
 =-\frac{2r^2}{3\beta^2}(\tilde m\beta)^3,
 \qquad {\rm thus,}\qquad
 -\frac12\log Z_1^{(0)}
 =\frac{r^2}{3\beta^2}(\tilde m\beta)^3.
 \label{eq:Z1-p0}
\end{equation}
\\

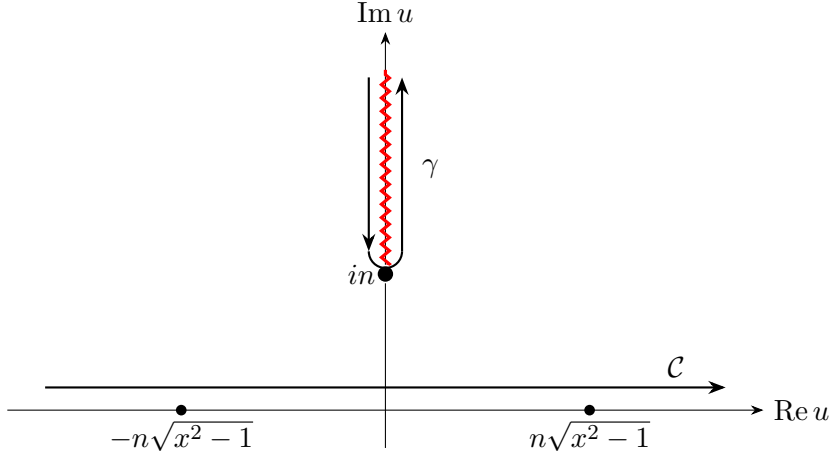
\begin{figure}[h]
	\centering
	\begin{tikzpicture}[>=Stealth]
		
		\draw[->] (-5,0) -- (5,0) node[right] {$\operatorname{Re}u$};
		\draw[->] (0,-0.5) -- (0,5) node[above] {$\operatorname{Im}u$};
		
		\fill (-2.7,0) circle (2pt);
		\fill ( 2.7,0) circle (2pt);
		
		\node[below] at (-2.7,0) {$-n\sqrt{x^2-1}$};
		\node[below] at ( 2.7,0) {$n\sqrt{x^2-1}$};
		
		\draw[red, very thick, decorate,
		decoration={zigzag, amplitude=1.5pt, segment length=5pt}]
		(0,1.8) -- (0,4.5);
		
		\filldraw[black, draw=white, line width=0.6pt] (0,1.8) circle (3.2pt);
		\node[left] at (0,1.8) {$in$};
		
		\draw[thick,->] (-4.5,0.3) -- (4.5,0.3);
		\node[above right] at (3.6,0.3) {$\mathcal{C}$};
		
		\draw[thick,->] (-0.22,4.4) -- (-0.22,2.1);
		\draw[thick] (-0.22,2.1)
		arc[start angle=180,end angle=360,radius=0.22];
		\draw[thick,->] (0.22,2.1) -- (0.22,4.4);
		
		\node[right] at (0.35,3.2) {$\gamma$};
		
	\end{tikzpicture}
	
	\caption{The contribution of the original contour $\cal C$ is equal to that of the contour $\gamma$ wrapping the branch cut, except the pole contribution at one end of the  branch cut at $u=in$.}
		\label{fig1}
\end{figure}

Now we will compute the term $\log Z_2(\tilde{m},\beta,\mu)$ from \eqref{eq:Z2-def} as a systematic series expansion in small $\beta/r$, following a similar method used in \cite{David:2024pir}. First we re-express this term as an integral representation as given below,
\begin{align}
 \log Z_2
 ={}&\frac{1}{2\pi}
 \sum_{l=0}^{\infty}\sum_{n=1}^{\infty}\frac{1}{n^2}
 \int_0^{\infty}\frac{\dd\tau}{\tau^2}
 \exp\!\left[-\tau n^2(\tilde m\beta)^2-\frac{1}{4\tau}\right]
 \nonumber\\
 &\times\int_{-\infty}^{\infty}\dd u\,
 \exp\!\left[-\frac{u^2}{4\tau n^2}
 -i\left(l+\frac12\right)\frac{\beta u}{r}\right]
 \sum_{m=-l}^{l}\e^{i\mu mn\beta}.
 \label{eq:Z2-schwinger}
\end{align}
Note the above integral representation linearizes the $l$ dependence in the exponential,  thus we can proceed to perform the sum over $m$ and $l$.
The sum over $m$ is straightforwardly carried out as a finite geometric sum, as follows
\begin{equation}
 \sum_{m=-l}^{l}\e^{i\mu mn\beta}
 =\cos(\beta l\mu n)
 +\cot\!\left(\frac{\beta\mu n}{2}\right)
 \sin(\beta l\mu n).
 \label{eq:interacting-msum}
\end{equation}
Now taking only the $l$-dependent part from \eqref{eq:Z2-schwinger} into account  and substituting \eqref{eq:interacting-msum} in it, we carry out the $l$-sum  as a sum of infinite geometric series in the following, leading to an exact compact expression
%
\begin{align}
 \sum_{l=0}^{\infty}
 \e^{-i\beta(l+\frac12)u/r}
 \sum_{m=-l}^{l} e^{i\mu mn\beta}& =
 \sum_{l=0}^{\infty}
 \e^{-i\beta(l+\frac12)u/r}
 \left[
 \cos(\beta l\mu n)
 +\cot\!\left(\frac{\beta\mu n}{2}\right)
 \sin(\beta l\mu n)
 \right]\nonumber,\\
 &=
 \frac{2\cos(\beta u/2r)}
 {(\e^{i\beta u/r}-\e^{i\beta\mu n})
  (1-\e^{-i\beta(\mu nr+u)/r})}\nonumber,\\
  &
 =
 \frac{2r^2}{\beta^2(\mu^2n^2r^2-u^2)}
 +\frac1{12}\left(
 \frac{u^2}{\mu^2n^2r^2-u^2}+2
 \right)
 +O\!\left((\beta/r)^2\right).
 \label{eq:Sn-highT}
\end{align}
At the last line of the above equation we have expanded the exact answer of this sum obtained in the 2nd line as a series expansion in small $\beta/r$.
Now we combine \eqref{eq:Sn-highT} with the equation \eqref{eq:Z2-schwinger} after performing the integral in $\tau$ in terms of modified Bessel functions of 2nd kind, leading to the following expression
\begin{align}
 \log Z_2
 ={}&\frac1{2\pi}\sum_{n=1}^{\infty}
 \int_{-\infty}^{\infty}\dd u\,
 \frac{4\tilde m\beta\,
 K_1\!\left(\tilde m\beta\sqrt{n^2+u^2}\right)}
 {\sqrt{n^2+u^2}}
 \nonumber\\
 &\times\Bigg[
 \frac{2r^2}{\beta^2(\mu^2n^2r^2-u^2)}
 +\frac1{12}\left(
 \frac{u^2}{\mu^2n^2r^2-u^2}+2
 \right)
 +O\!\left((\beta/r)^2\right)
 \Bigg].
 \label{eq:Z2-highT}
\end{align}
This provides a formal series expansion of $\log Z_2(\tilde m, \beta,\mu)$ in small $\beta/r$. 
From the above series expansion let us only focus on the leading high temperature contribution, as written below with the use of the definition $x^2=1+\mu^2r^2 $
\begin{equation}
 \log Z_{2}^{(0)}
 =\frac{8\tilde m r^2}{\beta}\sum_{n=1}^{\infty}
 \int_{-\infty}^{\infty}\frac{\dd u}{2\pi}
 \frac{K_1\!\left(\tilde m\beta\sqrt{n^2+u^2}\right)}
 {\sqrt{n^2+u^2}\,[(x^2-1)n^2-u^2]}.
 \label{eq:Z2-full-line}
\end{equation}
Note that  the integration contour in the above equation is defined slightly above the real line in the complex $u$-plane to avoid the poles of the integrand on the real line at $u^2=n^2(x^2-1)$.
The integrand admits a branch cut along $u\in (in,\infty)$ on the imaginary $u$-axis in the upper half plane, accompanied by a pole coincident on the branch point at $u=in$\footnote{The contribution due to the pole located at an endpoint of a branch cut discontinuity is fixed by the fact that the expression has to reproduce the free theory answer correctly when $\tilde m\to 0$, as there is no universal prescription given to handle such a pole.}. Thus the integral along the contour slightly above the real axis can be evaluated by correctly accounting for the branch cut and the pole contribution on the imaginary $u$ axis as shown in the \figref{fig1}.
Performing this manipulation we obtain the expression below with the use of the definition $q=\sqrt{y^2-n^2}$.

\begin{equation}
 \log Z_2^{(0)}
 =\frac{4r^2}{\beta^2x^2}\zeta(3)
 -\frac{4\tilde m r^2}{\beta}\sum_{n=1}^{\infty}\int_0^{\infty}\dd q\,
 \frac{J_1(\tilde m\beta q)}{(q^2+n^2x^2)\sqrt{q^2+n^2}}.
 \label{eq:contour-rotated}
\end{equation}

The first term in the above equation is independent of $\tilde m$, it comes as the contribution from the pole and the 2nd term arises due to the branch cut discontinuity.
As an internal consistency check, one can notice that at $\tilde m\to0$ only the first term from the above equation survives and the complete $\log Z_1$ contribution from \eqref{eq:Z1-highT} vanishes as well, therefore when combined with \eqref{eq:interacting-split} correctly reproduces the free theory result \eqref{eq:free-leading-pole}.\\


Now, the second term in \eqref{eq:contour-rotated} can be rewritten using a
Feynman parametrization\footnote{We have used the integral formula,
$\displaystyle
 \frac{1}{A\sqrt B}
 =\frac12\int_0^1\frac{\dd v}{\sqrt{1-v}}
 \frac{1}{[vA+(1-v)B]^{3/2}}$ for $A,B>0$, to rewrite the denominator of the integrand in the first line of \eqref{eq:branch-feynman-parametrized}. } which will allow us to carry out the sum over $n$ as given below 
\begin{align}
 \log Z_{2,{\rm mass}}^{(0)}
 &\equiv-\frac{4\tilde m r^2}{\beta}
 \sum_{n=1}^{\infty}\int_0^{\infty}\dd q\,
 \frac{J_1(\tilde m\beta q)}{(q^2+n^2x^2)\sqrt{q^2+n^2}}
 \nonumber,\\
 &=-\frac{2\tilde m r^2}{\beta}
 \sum_{n=1}^{\infty}\int_0^1\frac{\dd v}{\sqrt{1-v}}
 \int_0^\infty\dd q\,
 \frac{J_1(\tilde m\beta q)}
 {[q^2+n^2(1-v+vx^2)]^{3/2}}.
 \label{eq:branch-feynman-parametrized}
\end{align}
Now one can perform the integral over $q$ in the above expression using the standard integral formula of the Bessel function as given below
\begin{equation}
 \int_0^\infty\dd q\,
 \frac{J_1(\tilde m\beta q)}
 {[q^2+n^2(1-v+vx^2)]^{3/2}}
 =\frac{
  1-\bigl[1+n\tilde m\beta\sqrt{1-v+vx^2}\bigr]
  \e^{-n\tilde m\beta\sqrt{1-v+vx^2}}
 }{\tilde m\beta\,n^3(1-v+vx^2)^{3/2}}.
 \label{eq:J1-integral}
\end{equation}
\\

Substituting \eqref{eq:J1-integral} in
\eqref{eq:branch-feynman-parametrized}, we perform the sum over $n$ in terms of Polylogarithm functions,
and using the change of variable $z=\sqrt{1-v+vx^2}$, we have
\begin{align}
 &\log Z_2^{(0)}(\tilde m\beta,x)
 =\frac{4r^2}{\beta^2}\left[
 \frac{\zeta(3)}{x^2}
 -\frac{1}{\sqrt{1-x^2}}
 \int_x^1\dd z\,
 \frac{\Phi(\tilde m\beta z)}{z^2\sqrt{z^2-x^2}}
 \right],\nonumber\\
 \label{eq:Z2-reduced}
& {\rm with} \qquad \Phi(w)\equiv
 \zeta(3)-\Li_3(e^{-w})-w\Li_2(e^{-w}).
\end{align}
Now combining \eqref{eq:Z2-reduced}, \eqref{eq:Z1-p0} and \eqref{eq:interacting-split} we obtain the leading high temperature contribution to the free energy per scalar component given as follows
%
\begin{equation}
 \frac{\beta^2}{r^2}\log Z^{(0)}(\tilde m\beta,x)
 =\frac{(\tilde m\beta)^3}{3}+\frac{2\zeta(3)}{x^2}
 -\frac{2}{\sqrt{1-x^2}}
 \int_x^1\dd z\,
 \frac{\Phi(\tilde m\beta z)}{z^2\sqrt{z^2-x^2}}.
 \label{eq:physical-functional}
\end{equation}
Thus we have reduced the leading high temperature contribution to the free energy  into a form which no longer involves any infinite sum, only an integral remains to be performed. This integral is convergent and can be computed numerically, however we could not find any analytic closed form expression. We will see that 
this representation proves particularly useful,  as it allows us to obtain systematic series expansions of $\log Z^{(0)}(\tilde m \beta,x)$ both around $x=0$ and $x^2=1$. Before proceeding to computations of such expansions, we will evaluate the gap equation by differentiating the leading high temperature expression for the free energy obtained here with respect to $\tilde m$.
\subsection{Gap equation at high temperature}
\label{sec:gap}

Differentiating the mass-dependent terms in \eqref{eq:physical-functional} and setting it to zero gives the saddle point condition as was given in \eqref{eq:critical-saddle-general}
\begin{equation}
 \frac{\partial}{\partial(\tilde m\beta)}\left(\frac{\beta^2}{r^2}\log Z^{(0)}\right)
 =\tilde m\beta\left[
 \tilde m\beta+\frac{2}{\sqrt{1-x^2}}
 \int_x^1\dd z\,
 \frac{\log(1-\e^{-\tilde m\beta z})}{\sqrt{z^2-x^2}}
 \right].
 \label{eq:dF-dmass}
\end{equation}
For compactness, we define the integral appearing  in \eqref{eq:dF-dmass} to be $I(x,\tilde m \beta )$, as given below
\begin{equation}
 I(x,\tilde m\beta)=\int_x^1\dd z\,
 \frac{\log(1-\e^{-\tilde m\beta z})}{\sqrt{z^2-x^2}}.
 \label{eq:I-finite}
\end{equation}
Substituting the definition \eqref{eq:I-finite} into \eqref{eq:dF-dmass}, the non-trivial 
gap equation becomes
\begin{equation}
 \tilde m\beta=-\frac{2}{\sqrt{1-x^2}}I(x,\tilde m\beta).
 \label{eq:gap-I}
\end{equation}
This equation gives the non-trivial gap equation for the thermal mass in presence of the angular potential, parametrized by $x$, at the leading order in high temperature. This gap equation can be solved numerically as shown in  \figref{fig:thermal-mass-notebook}. In the remainder of this section we will solve the gap equation near $x^2=1$ as well as $x=0$ as systematic order by order expansions, thereby evaluating the series expansions for free energies at these limits.

\subsection{Small $\mu$ expansion}
\label{subsec:small-mu}
In this subsection we expand the gap equation \eqref{eq:gap-I} about
$x^2=1$, which corresponds to $\mu=0$ by the definition
$x^2=1+\mu^2r^2$, and determine the thermal mass order by order in
$1-x^2$.  We use the substitution
$z=\sqrt{x^2+(1-x^2)t^2}$ in \eqref{eq:gap-I} so that the integration range becomes independent of $x$
and  all dependences on $x$ are absorbed in the integrand.  The gap equation then
takes the form
\begin{align}
\tilde m\beta+2\int_0^1\dd t\,
 \frac{\log \bigl(1-\e^{-\tilde m\beta
 \sqrt{x^2+(1-x^2)t^2}}\bigr)}
 {\sqrt{x^2+(1-x^2)t^2}}=0.
 \label{eq:gap-fixed}
\end{align}
Applying the same change of variable to the free energy
\eqref{eq:physical-functional} gives
\begin{align}
 \frac{\beta^2}{r^2}\log Z^{(0)}(\tilde m\beta,x)
 ={}&\frac{\left(\tilde m\beta\right)^3}{3}+\frac{2\zeta(3)}{x^2}
 -2\int_0^1\dd t\,
 \frac{\Phi\big(\tilde m\beta\sqrt{x^2+(1-x^2)t^2}\big)}
 {[x^2+(1-x^2)t^2]^{3/2}}.
 \label{eq:F-fixed}
\end{align}
Now we expand the integrand in the 2nd terms of the gap equation \eqref{eq:gap-fixed} around $x^2=1$ and that can be written as the following general expression
%
\begin{align}
 \frac{\log\bigl(1-\e^{-\tilde m\beta
 \sqrt{x^2+(1-x^2)t^2}}\bigr)}
 {\sqrt{x^2+(1-x^2)t^2}}
 =-\sum_{n=0}^{\infty}
 \frac{[(1-x^2)(1-t^2)]^n}{n!}
 \sum_{j=0}^{n}
 \frac{(2n-j)!(\tilde m\beta)^j}{2^{2n-j}(n-j)!j!}
 \Li_{1-j}(\e^{-\tilde m\beta}).
 \label{eq:gap-kernel-derivatives}
\end{align}
Now we substitute \eqref{eq:gap-kernel-derivatives} into \eqref{eq:gap-fixed}, and carry out the integral over $t$ straightforwardly to obtain the following gap equation expanded in systematic series expansion about $x^2=1$.
\begin{align}
\tilde m\beta
 -2\sum_{n=0}^{\infty}
 \frac{n!(1-x^2)^n}{(2n+1)!}
 \sum_{j=0}^{n}
 \frac{2^j(2n-j)!}{(n-j)!j!}
 (\tilde m\beta)^j\Li_{1-j}(\e^{-\tilde m\beta})=0.
 \label{eq:gap-x1-all-orders}
\end{align}
Note that the leading order gap equation correctly reproduces the gap equation in absence of the angular potential well known in the literature \cite{Sachdev:1993pr,Chubukov:1993aau}
\begin{align}
 &\tilde m_0\beta+2\log(1-\e^{-\tilde m_0\beta})=0,
 \nonumber\\
&{\rm thus}\qquad \tilde m_0\beta
 =2\log\!\Big(\frac{1+\sqrt5}{2}\Big).
 \label{eq:m0-value}
\end{align}
For compactness, we denote the golden ratio by
$\rho_{\rm g}\equiv(1+\sqrt5)/2$.  Now solving the gap equation
\eqref{eq:gap-x1-all-orders} iteratively in powers of $1-x^2$ gives the
following thermal-mass expansion:
\begin{align}
 \frac{\tilde m(x)\beta}{2\log\rho_{\rm g}}
 =&1+\frac{1-x^2}{3}
 +\frac{45+2\sqrt5\log\rho_{\rm g}}{225}(1-x^2)^2
 \nonumber\\
 &+\frac{2025+186\sqrt5\log\rho_{\rm g}-40\log^2\rho_{\rm g}}{14175}
 (1-x^2)^3
 \nonumber\\
 &\hspace{-1 cm}+\frac{118125-5560\log^2\rho_{\rm g}
 +\sqrt5\log\rho_{\rm g}(16020+378\log^2\rho_{\rm g})}{1063125}
(1-x^2)^4
 +O((1-x^2)^5).
 \label{eq:mass-x1-series}
\end{align}
The agreement of the above solution with the numerical solution of \eqref{eq:gap-fixed} near $x^2=1$ is shown in figure \ref{fig:thermal-mass-notebook}.
Similarly we expand the expression for the free energy given in \eqref{eq:F-fixed} as a systematic expansion around  $x^2=1$, then we substitute the expression for the thermal mass \eqref{eq:mass-x1-series} in it. This leads to the following expansion for the free energy at $x^2=1$.
%
%
\begin{align}
 &\frac{\beta^2}{r^2}\log Z_{\rm saddle}^{(0)}(x)
 =\frac85\zeta(3)+\frac85\zeta(3)(1-x^2)
+\Big(\frac85\zeta(3)
 +\frac{4\sqrt5\log^3\rho_{\rm g}}{45}\Big)(1-x^2)^2
 \nonumber\\
 &\qquad\qquad+\Big(\frac85\zeta(3)
 +\frac{4\log^3\rho_{\rm g}(127\sqrt5-8\log\rho_{\rm g})}{2835}\Big)
 (1-x^2)^3
 \nonumber\\
 &+\left[\frac85\zeta(3)
 +\frac{4\log^3\rho_{\rm g}}{70875}
 \left(\sqrt5(4590+22\log^2\rho_{\rm g})
 -435\log\rho_{\rm g}\right)\right]
 (1-x^2)^4
 +O((1-x^2)^5).
 \label{eq:F-x1-final}
\end{align}
The leading term reproduces the free energy at the non-trivial fixed point in absence of the angular potential at high temperature.  The first correction term arising due to the angular potential has also been computed in  \cite{Mauro:2026zus}. We have tested this expansion\footnote{
	The expansions for both the thermal mass and the free energy are also found numerically till $O((1-x^2)^{12})$, available in the ancillary file \texttt{x1\_series.txt} with the arXiv version of this paper.} by evaluating the free energy  with the use of the numerical solution of the gap equation \eqref{eq:gap-I} in \figref{fig:logZ-notebook}, demonstrating the agreement between them in the region near $x^2=1$
.
\subsection{Expansion at $\mu^2r^2=-1$}
\subsubsection{Gap equation and its solution}
We now solve the gap equation near $\mu^2r^2=-1$, equivalently $x\to0$ using the definition used earlier $x^2=1+\mu^2r^2$.
Before proceeding to the detailed analysis, we first examine the gap equation \eqref{eq:gap-I} qualitatively to identify the plausible asymptotic behavior of the thermal mass as a consistent solution of the 
%
 the gap equation at $x\to 0$.

We recall the definition of the integral $I(x,\tilde m\beta)$ from \eqref{eq:I-finite} involved in the gap equation \eqref{eq:gap-I}
\begin{align}\label{I integral}
 I(x,\tilde m\beta)=\int_x^1\dd z\,
 \frac{\log(1-\e^{-\tilde m\beta z})}{\sqrt{z^2-x^2}}.
\end{align}
The above  integral admits logarithmic divergence as $x\to 0$ which arises from the expansion of the numerator of the integrand in small $z$ whose leading behavior is noted below
\begin{equation}
	\log(1-\e^{-\tilde m\beta z})
	=\log(\tilde m\beta z)+O(\tilde m \beta z).
	\label{eq:bernoulli-expansion1}
\end{equation}
This leads to a double logarithmic growth of $I(x,\tilde m\beta)$ at $x\to0$, as demonstrated below
%
\begin{equation}
 I(x,\tilde m\beta)\sim
 -\frac12\log^2\!\frac1x.
\end{equation}
Now incorporating this leading behavior of $I(x,\tilde m\beta )$ in the gap equation \eqref{eq:gap-I}, one obtains the following relation
\begin{align}\label{sim gap}
	\tilde m\beta\sim
	\log^2\!\frac1x.
\end{align}
Hence the qualitative regime of behavior  where such relation holds is given by 
\begin{equation}
 \tilde m\beta\longrightarrow\infty,
 \qquad
 \tilde m\beta x\longrightarrow0\qquad{\rm as }\qquad x\longrightarrow 0.
 \label{eq:double-scaling}
\end{equation}
More precisely we can get the behavior of the $\tilde m\beta $ that solves the relation \eqref{sim gap}, at the leading order in small $x$
\begin{align}\label{mb mbx}
\tilde m\beta\sim{}&\log^2\!\frac1x \qquad {\rm and } \qquad\tilde m \beta x\sim x \log^2\!\frac1x .
\end{align}
Similarly one can also estimate the leading behavior of the integral involved in the free energy  \eqref{eq:physical-functional}, followed by substitution of the thermal mass as given in \eqref{mb mbx}, to obtain the leading small $x$ behavior of the free energy as  
\begin{align}
	 \frac{\beta^2}{r^2}\log Z_{\rm saddle}^{(0)}(x)
	\sim{}&\frac{2\zeta(3)}{x^2}
	-\frac16\log^6\!\frac1x .
\end{align}
Here we note that logarithmic behavior arises due to the leading solution for the thermal mass, while first term is the pole contribution identical to that present in the free theory.
\\

In principle one should be able to understand the perturbative structure for the gap equation by 
substituting the higher-order expansion \eqref{eq:bernoulli-expansion1} into \eqref{I integral}, followed by term-by-term integration. However, organizing the resulting terms systematically in a manner consistent with the leading behavior \eqref{sim gap} becomes cumbersome.
But the qualitative behavior \eqref{eq:double-scaling} leads to the following statement. A perturbative expansion of the gap equation in small $\tilde m \beta x$, can prove the consistency of  this leading solution and allows computations of systematic corrections. In the remainder of this section we will demonstrate that  such a perturbative expansion for the gap equation indeed exists, following a more careful treatment.
\\

Now to obtain the complete systematic expansion of the gap equation, we use a different approach from what was used in estimating  the leading qualitative behavior above. First we rewrite the integral $I(x,\tilde m\beta)$ from \eqref{eq:I-finite} in the following manner
%
\begin{equation}
	I(x,\tilde m\beta)=I_\infty(\tilde m\beta x)
	-\int_1^\infty\dd z\,
	\frac{\log(1-\e^{-\tilde m\beta z})}{\sqrt{z^2-x^2}}.
	\label{eq:I-tail-exact}
\end{equation}
where
\begin{equation}
	I_\infty(\tilde m\beta x)=\int_x^\infty\dd z\,
	\frac{\log(1-\e^{-\tilde m\beta z})}{\sqrt{z^2-x^2}}.
	\label{eq:I-infty}
\end{equation}
The second term in \eqref{eq:I-tail-exact} is positive and exponentially suppressed at large $\tilde m \beta$. We  denote this term by
$I_{\rm tail}$.  Its large-$\tilde m\beta$ behavior follows directly by
using the exponential expansion of
$\log(1-\e^{-\tilde m\beta z})$ and expanding  $\sqrt{z^2-x^2}$ in the denominator about $z=1$, and finally integrating term by term. This gives
\begin{align}
	I_{\rm tail}(x,\tilde m\beta)
	={}&\frac{\e^{-\tilde m\beta}}{\tilde m\beta}
	\Bigg[
	\frac1{\sqrt{1-x^2}}
	-\frac1{\tilde m\beta(1-x^2)^{3/2}}
	+\frac{2+x^2}{(\tilde m\beta)^2(1-x^2)^{5/2}}
	+O\!\left((\tilde m\beta)^{-3}\right)
	\Bigg]\nonumber\\
	&\qquad\qquad\qquad\qquad\qquad\qquad\qquad\qquad\qquad\qquad\qquad +O\!\left(\frac{\e^{-2\tilde m\beta}}{\tilde m\beta}\right).
	\label{eq:I-tail-asymptotic}
\end{align}
Thus  combining \eqref{eq:I-tail-exact} with
\eqref{eq:I-tail-asymptotic} gives
\begin{equation}
	I(x,\tilde m\beta)=I_\infty(\tilde m\beta x)
	+O\!\left(\frac{\e^{-\tilde m\beta}}{\tilde m\beta}\right).
	\label{eq:I-tail-estimate}
\end{equation}
Now we perform the integral over $z$ in $I_\infty$ defined in \eqref{eq:I-infty} by expanding the logarithm in the numerator of the integrand in small $e^{-\tilde m \beta z}$. This results into a series involving modified Bessel functions of 2nd kind, as shown below 
\begin{align}
	-I_\infty(\tilde m\beta x)
	&=\sum_{n=1}^{\infty} \frac{1}{n} \int_x^\infty \frac{e^{-n\tilde m  \beta z}}{\sqrt{z^2-x^2}} dz
	\nonumber,\\
	&=
	\sum_{n=1}^{\infty}
	\frac{K_0(n\tilde m\beta x)}{n}.
	\label{eq:BP-bessel-sum}
\end{align}
The next task is to obtain an expansion for $I_\infty$ consistent with the leading behavior qualitatively  discussed before.
We avoid direct small $\tilde m \beta x$ expansion of  Bessel functions in the above expression as summing over $n$ is not straightforward after that. But the above sum of the Bessel functions can be reorganized as the following expansion with the use of the Poisson resummation guided by a set of algebraic manipulations shown in \appref{app:bessel-poisson}.
%
%
\begin{align}
	-I_\infty(\tilde m\beta x)=&
 \frac12\log^2\!\frac2{\tilde m\beta x}+C_0
-\sum_{k=1}^{\infty}(\tilde m\beta x)^{2k}
 \Big[
 \frac{B_{2k}}{2k\,4^k(k!)^2}
 \log\!\frac2{\tilde m\beta x}+C_{2k}
 \Big],
\nonumber\\
{\rm where}\qquad &C_0
 =\frac{\pi^2}{24}-\gamma_1-\frac{\gamma_E^2}{2}, \label{eq:I-all-orders}
\\ {\rm and} \qquad
& C_{2k}
 =\frac{B_{2k}}{2k\,4^k(k!)^2}
 \left[
 \log(2\pi)+H_k-H_{2k-1}
 -\frac{\zeta'(2k)}{\zeta(2k)}
 \right] \qquad {\rm for }\qquad k\ge1.
 \nonumber
\end{align}
Here $H_n=\sum_{j=1}^n j^{-1}$, with $H_0=0$ and $\gamma_E$ and $\gamma_1$ are the standard Stieltjes-constants.\footnote{The Stieltjes constants are defined by
	$
	\zeta(s)=\frac1{s-1}
	+\sum_{n=0}^{\infty}\frac{(-1)^n\gamma_n}{n!}(s-1)^n
	$, with $\gamma_E=\gamma_0.$}
\\

\begin{figure}[tbp]
 \centering
 \includegraphics[width=0.88\textwidth]{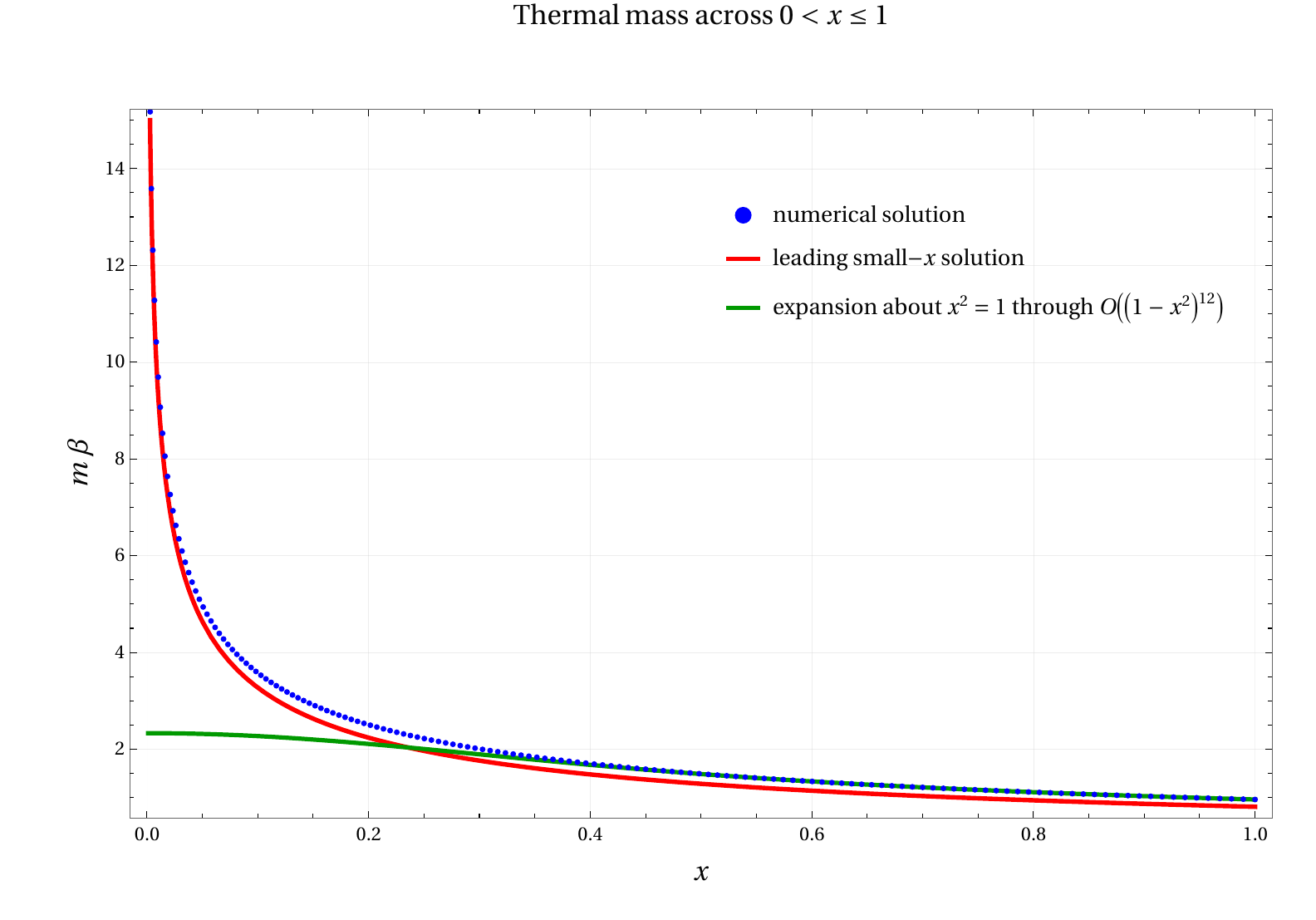}
 \caption{Dimensionless thermal mass $\tilde m\beta$ on
 $10^{-3}\leq x\leq1$.  The blue points are the  numerical solutions
 of \eqref{eq:gap-fixed}; the red curve is the leading small-$x$ solution
 \eqref{Prodlog} given by the Productlog function, and the green curve is the expansion
 \eqref{eq:mass-x1-series} about $x=1$, continued through order
 $(1-x^2)^{12}$ available in the ancillary file with the arXiv version of this paper. }
 \label{fig:thermal-mass-notebook}
\end{figure}

Finally, substituting \eqref{eq:I-all-orders} into \eqref{eq:gap-I} and using
\eqref{eq:I-tail-estimate} gives
\begin{align}
 \tilde m\beta={}&\frac1{\sqrt{1-x^2}}
 \Bigg[
 \log^2\!\frac2{\tilde m\beta x}+2C_0
  -2\sum_{k=1}^{\infty}(\tilde m\beta x)^{2k}
 \left(
 \frac{B_{2k}}{2k\,4^k(k!)^2}\log\!\frac2{\tilde m\beta x}+C_{2k}
 \right)
 \Bigg]
 \nonumber\\
 &\qquad\qquad\qquad\qquad+O\!\left(
 \frac{\e^{-\tilde m\beta}}
 {\tilde m\beta\sqrt{1-x^2}}\right).
 \label{eq:gap-expanded}
\end{align}
Thus we have obtained a systematic series expansion of the gap equation \eqref{eq:gap-I} in small $\tilde m \beta x$ consistent with the leading asymptotics provided in \eqref{eq:double-scaling} and \eqref{mb mbx}. Now we can solve the leading order gap equation as shown below
\begin{equation}
 \tilde m\beta=\log^2\!\frac{2}{\tilde m\beta x}.
 \label{eq:mass-leading-W}
\end{equation}
This leading order gap equation admits a solution in terms of Productlog function as the following
\begin{align}\label{Prodlog}
	\tilde m\beta=4W_0^2\!\left(\frac1{\sqrt{2x}}\right)\sim  \log^2\!\frac1x ,
\end{align}
where $W_0(x)$ denotes the Productlog function of $x$. This leading solution in terms of the Productlog function agrees with the numerical solution of the gap equation \eqref{eq:gap-I} near $x=0$ as demonstrated in the figure \ref{fig:thermal-mass-notebook}. 
The thermal mass is evaluated till a few higher orders by solving the gap equation \eqref{eq:gap-expanded} systematically order by order in \appref{app:thermal-mass-orders}, leading  to the following   answer
%
\begin{align}
 \tilde m\beta={}&\log^2\!\frac1x
 -4\log\!\frac1x\,\log\log\!\frac1x
 +2\log2\,\log\!\frac1x
 +4\log^2\log\!\frac1x
 \nonumber\\
 &+(8-4\log2)\log\log\!\frac1x
 +\log^22-4\log2+2C_0
 \nonumber\\
 &+O\!\left(
 \frac{\log^2\log(1/x)}{\log(1/x)}\right)
 +O\!\left(x^2\log^5\!\frac1x\right).
 \label{eq:mass-corrected}
\end{align}
Such a solution has
$\tilde m\beta\to\infty$ and $\tilde m\beta x\to0$, thereby verifying
\eqref{eq:double-scaling}. Note that $C_0$ is provided in \eqref{eq:I-all-orders}.

\subsubsection{$\log Z$ as a small-$x$ expansion}
\label{sec:logZ}
In this subsection we will evaluate the free energy \eqref{eq:physical-functional} at the value of the thermal mass \eqref{eq:mass-corrected} satisfying the gap equation \eqref{eq:gap-expanded}. First we will expand the free energy as a  series expansion  in small $\tilde m \beta x$,  consistent with the qualitative behavior presented in \eqref{eq:double-scaling}. More precisely we have to obtain such a series expansion for the integral in the last term of \eqref{eq:physical-functional}. Thus let us consider only this term as follows
%
\begin{equation}
 \frac12\log Z_{2,{\rm mass}}^{(0)}
 =-\frac{2r^2}{\beta^2\sqrt{1-x^2}}
 \int_x^1\dd z\,
 \frac{\Phi(\tilde m\beta z)}{z^2\sqrt{z^2-x^2}}.
 \label{eq:Zth-z}
\end{equation}
Now the above integral is related to the integral $I(x,\tilde m\beta)$  in the following manner
\begin{equation}
 \frac12\log Z_{2,{\rm mass}}^{(0)}
 =\frac{2r^2}{\beta^2\sqrt{1-x^2}}
 \int_0^{\tilde m\beta}\lambda I(x,\lambda)\,\dd\lambda.
 \label{eq:Zth-lambda}
\end{equation}
This follows from the fact that the integral $I(x,\tilde m \beta)$ was originated from the differentiation of the term \eqref{eq:Zth-z} in the free energy \eqref{eq:physical-functional}, as was shown in \eqref{eq:dF-dmass}, up to an overall factor of $\tilde m$, apart from other $\tilde m$-independent factors.\\

Now using \eqref{eq:I-tail-exact} in the above equation we have 
\begin{align}
	\frac{1}{2}\log Z_{2,{\rm mass}}^{(0)}=\frac{2r^2}{\beta^2\sqrt{1-x^2}}
	\int_0^{\tilde m\beta}\lambda \Big[I_\infty(x\lambda)-\int_1^\infty\dd z\,
	\frac{\log(1-\e^{-\lambda z})}{\sqrt{z^2-x^2}}\Big]\,\dd\lambda.
	\label{eq:Zth-lambda1}
\end{align}
The 2nd term from the above expression is evaluated as follows
\begin{align}\label{2nd term}
&	-\frac{2r^2}{\beta^2\sqrt{1-x^2}}
	\int_0^{\tilde m\beta}d\lambda\,\lambda \int_1^\infty\dd z\,
	\frac{\log(1-\e^{-\lambda z})}{\sqrt{z^2-x^2}}\nonumber\\
	&=\frac{2r^2}{\beta^2\sqrt{1-x^2}}
	\int_1^\infty\dd z\,
	\frac{\zeta (3)-\tilde m\beta z \text{Li}_2\left(e^{-\tilde m\beta z}\right)-\text{Li}_3\left(e^{-\tilde m\beta z}\right)}{z^2 \sqrt{z^2-x^2}}\nonumber,\\
	&=\frac{2r^2 \zeta (3)}{\beta^2\sqrt{1-x^2} (1+\sqrt{1-x^2})}+O(e^{-\tilde{m}\beta}).
\end{align}
One can easily realize that the contribution due to the Polylogarithm functions in the numerator of the integrand in the 2nd line of the above equation, are exponentially suppressed at large $\tilde m\beta$, following a similar treatment as was used in \eqref{eq:I-tail-asymptotic}.
\\

\begin{figure}[tbp]
 \centering
 \includegraphics[width=0.88\textwidth]{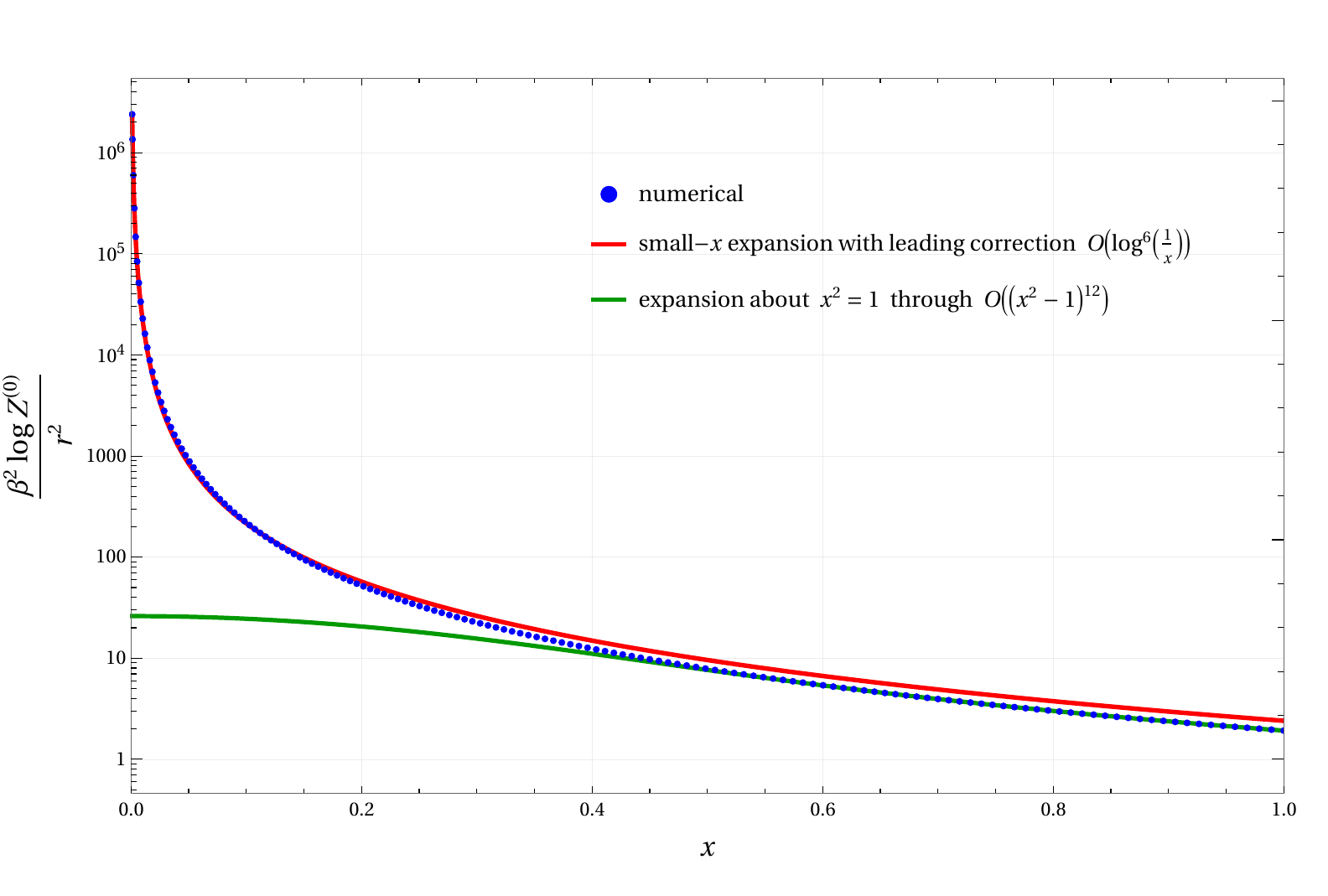}
 \caption{The leading high-temperature partition function
 $\beta^2\log Z_{}^{(0)}/r^2$ on $10^{-3}\leq x\leq1$, displayed
 on a logarithmic vertical scale.  The blue points are obtained by numerically solving
 \eqref{eq:gap-fixed} and thereby evaluating \eqref{eq:F-fixed}; the red curve is
 the leading terms of the small-$x$ expansion \eqref{eq:F-small-x},
 $2\zeta(3)/x^2-\frac16\log^6(1/x)$, and the green curve is the expansion
 \eqref{eq:F-x1-final} about $x=1$, continued through order
 $(1-x^2)^{12}$, as given in the ancillary file with the arXiv version of this paper.}
 \label{fig:logZ-notebook}
\end{figure}

The integral in the first term in \eqref{eq:Zth-lambda1} is computed by straightforward term by term integration of the expansion for $I_{\infty}$ given in \eqref{eq:I-all-orders}
\begin{align}
\int_0^{\tilde m\beta }\lambda I_\infty(x,\lambda)\,\dd \lambda
 ={}&-\frac{\left(\tilde m\beta\right)^2}{4}
 \left[\log^2\!\frac2{\tilde m\beta x}+\log\!\frac2{\tilde m\beta x}+\frac12\right]
 -\frac{C_0\left(\tilde m\beta\right)^2}{2}
 \nonumber\\
 &+\left(\tilde m\beta\right)^2\sum_{k=1}^{\infty}\frac{(\tilde m\beta x)^{2k}}{2k+2}
 \left[
 \frac{B_{2k}}{2k\,4^k(k!)^2}
 \left(\log\!\frac2{\tilde m\beta x}+\frac1{2k+2}\right)+C_{2k}
 \right].
 \label{eq:int-I-result}
\end{align}
Combining \eqref{eq:Zth-lambda1}, \eqref{2nd term} and \eqref{eq:int-I-result}, 
we have the following series expansion at $x=0$
we obtain the free energy at the non-trivial fixed point at the high temperature as a systematic series expansion at $x=0$ 
\begin{align}
 &\frac{\beta^2}{2r^2}\log Z_{2,{\rm mass}}^{(0)}
 =\frac{2\zeta(3)}{\sqrt{1-x^2}[1+\sqrt{1-x^2}]}
-\frac{\left(\tilde m\beta\right)^2}{2\sqrt{1-x^2}}
 \left[\log^2\!\frac2{\tilde m\beta x}+\log\!\frac2{\tilde m\beta x}+\frac12+2C_0\right]
 \nonumber\\
 &+\frac{2\left(\tilde m\beta\right)^2}{\sqrt{1-x^2}}
 \sum_{k=1}^{\infty}\frac{(\tilde m\beta x)^{2k}}{2k+2}
 \Bigg[
 \frac{B_{2k}}{2k\,4^k(k!)^2}
 \log\!\frac2{\tilde m\beta x}
 +\frac{B_{2k}}{2k\,4^k(k!)^2(2k+2)}
 +C_{2k}
 \Bigg]
 +O(\tilde m\beta\e^{-\tilde m\beta}).
 \label{eq:Zth-offshell}
\end{align}
Now using \eqref{eq:Zth-offshell} in the \eqref{eq:physical-functional} and finally substituting the series expansion for the thermal mass \eqref{eq:mass-corrected} in it, we find the free energy to be 
\begin{align}
 \frac{\beta^2}{r^2}\log Z_{\rm saddle}^{(0)}(x)
 =\frac{2\zeta(3)}{x^2}
 -\frac16\log^6\!\frac1x
 +\left(2\log\log\!\frac1x-\log2-\frac12\right)
 \log^5\!\frac1x
 \nonumber\\
 +\Big[
 -10\log^2\log\!\frac1x
 +(10\log2+1)\log\log\!\frac1x
-\frac52\log^22-\frac12\log2-\frac14-C_0
 \Big]\log^4\!\frac1x
 \nonumber\\
 +O\!\left(\log^3\!\frac1x\,\log^3\log\!\frac1x\right).
 \label{eq:F-small-x}
\end{align}
We demonstrate that a good agreement with the free energy \eqref{eq:physical-functional} evaluated at the numerical solution of the gap equation \eqref{eq:gap-I} is observed just by considering the leading correction to in the above expansion in the figure
\ref{fig:logZ-notebook}.

\section{The model on the pp-wave geometry}
\label{sec:plane-wave}

Recently in \cite{Komargodski:2026ain}, it was argued that the residue of the pole at  large angular velocity limit 
for conformal field theories can be obtained by studying the theory in the plane wave geometry. 
Let $x^2 = 1- \hat \mu^2 r^2 $, 
 the  partition function in the $x\rightarrow 0 $ limit  is then given by  \footnote{We have related the 
	$\epsilon $ in equation (3.12)  of \cite{Komargodski:2026ain} to $x$ used in this paper. The relation is given by $x^2 = 2\epsilon$.  }
\begin{eqnarray}
	\log Z( \beta , x) = \frac{4\pi   }{x^2}  \log Z_{pp} \Big( \frac{\beta}{r}  \Big) ,
\end{eqnarray}
$Z_{pp} \big( \frac{\beta}{r} \big)$ is the partition function  on the conformal field theory on the pp wave. 
In this section, we would like to study the free energy of the critical vector model at large $N$ described in \eqref{eq:zero-mode-path-integral}. 
This will enable us to obtain its partition function  $Z_{pp} \big( \frac{\beta}{r} \big) $ and allow us to 
check the consistency of the result obtained  from solving the gap equation  on $S^1\times S^2$ 
for all values of the angular velocity 
in section. 
To study the critical vector model on the pp wave background 
we follow the same procedure as was done on $S^1\times S^2$ in the section \ref{sec:interacting}. 
We  evaluate the partition function of the massive scalar  on the pp wave background and look for the 
saddle point by minimising the partition function with respect to mass to obtain the gap equation. 

We begin by  re-deriving  the plane wave metric to keep track  of the factors involving the radius of the sphere. 
Following the 
analysis in  \cite{Komargodski:2026ain}, 
we begin with the metric
\begin{eqnarray}
	ds^2 = - dt^2 + r^2 ( d\theta^2 + \sin^2 \theta d\phi^2) .
\end{eqnarray}
We then define 
\begin{eqnarray}
	\phi =\varphi + \frac{(1-\epsilon)}{r}t.
\end{eqnarray}
With this the angular velocity $r\dot\phi $ is close to the speed of light as $\epsilon\to0$. 
Substituting in the metric, we obtain 
\begin{eqnarray}
	ds^2 = - dt^2 + r^2 d\theta^2 + r^2 \sin^2 \theta 
	\Big( d\varphi +  \frac{(1-\epsilon)}{r}dt \Big)^2 .
\end{eqnarray}
Now we scale the coordinates as follows
\begin{eqnarray}
	\varphi = \epsilon \,x , \qquad \theta = \frac{\pi}{2} + \sqrt{2\epsilon}\, y . 
\end{eqnarray}
Substituting this  in the metric and taking $\epsilon\rightarrow 0$, the leading terms in the metric organise as 
\begin{eqnarray}
	ds^2 = 2 \epsilon \Big( -( 1+ y^2 ) dt^2 + r^2 dy^2 + r dx dt\Big) .
\end{eqnarray}
Note, all dimensions are consistent. $y, \epsilon, x $ are dimensionless.
After a Weyl rescaling we obtain the metric of the form 
\begin{eqnarray} \label{ppwave}
	ds^2 = \Big( -( 1+ y^2 ) dt^2 + r^2 dy^2 + r dx dt\Big) .
\end{eqnarray}

The next step  is to evaluate the partition function of a massive scalar on the geometry in (\ref{ppwave}). 
The Ricci curvature of the geometry is zero and therefore the action  of  massive scalar in this background 
is given by 
\begin{eqnarray}
	S_{\hat m } (\phi)  =\frac{1}{2} \int dt dx dy \sqrt{g} 
	\left[ g^{\mu\nu} \partial_\mu \phi \partial_\nu \phi + \hat m^2  \phi^2 \right].
\end{eqnarray}
Here $\hat m$ is the thermal mass arising from the zero mode of the auxiliary field in the Hubbard Stratonovich transformation of the action of the vector  model at large $N$ in the pp-wave geometry. Again, we take the field $\hat m$ to be uniform. 
In fact as we show in the appendix, the pp-wave geometry allows $\hat m$ to be non-uniform in the coordinate $y$  and therefore just 
as in the $S^1\times S^2$, this model is not equivalent to the standard $O(N)$ model with quartic interaction at large $N$. 

To evaluate the partition function on this pp-wave background,  we first analytically continue to Euclidean time 
$t\rightarrow -i \tau$  with $\tau \sim \tau + \beta$ 
and solve for the eigen spectrum of the Laplacian which is given by 
\begin{eqnarray} \label{eigen1}
	\Big( \frac{ 4 i }{r} \partial_\tau \partial_x  + \frac{4}{r^2} ( 1+ y^2)\partial_x^2 +  \frac{1}{r^2} \partial_y^2  - \hat m ^2 
	\Big)\phi = \frac\lambda{r^2} \phi.
\end{eqnarray}
Here $\lambda$ is the eigen value, which we need to write in terms of  quantum numbers,  more suitable to evaluate 
the partition function. 
Since the pp-wave background is invariant under translations in  $x$ we can expand the massive scalar as 
\begin{eqnarray}
	\phi( \tau, x, y) = \sum_{n =-\infty}^\infty
	\int_{0}^\infty \frac{d k}{2\pi } e^{i k x } e^{\frac{2\pi i n \tau }{\beta} }  \phi ( n, k , y )  + cc .
\end{eqnarray}
Here we have also decomposed the scalar in its Matsubara modes and considered $k>0 $ for the modes to be well 
behaved as in \cite{Gibbons:1975jb,Komargodski:2026ain}. 
Substituting this expansion in (\ref{eigen1}) we obtain  the following eigen value equation for $\lambda$ 
\begin{eqnarray}
	(  - 4i  k \frac{2\pi n }{ r \beta}  - 4 ( 1+ y^2 ) \frac{ k^2}{r^2}     +  \frac{1}{r^2} \partial_y^2  - \hat m^2   ) \phi(n, k  , y) = \frac{ \lambda}{r^2}  \phi( n, k,  y) .
\end{eqnarray}
Note that $\phi^*(n, k , y)$ also obeys the same equation. 
The $y$ dependence of the  Laplacian is that of the quantum mechanical harmonic oscillator, its eigen spectrum is given by 
\begin{equation}
	\frac{1}{r^2} \Big( - \partial_y^2 + 4  k^2 y^2 \Big) \psi_l(y) =  \Big( l + \frac{1}{2} \Big) \frac{4k }{r^2}  \psi_l(y), 
	\qquad l = 0, 1, 2, \cdots .
\end{equation}
Here $\psi_l(y)$ are the eigen functions of the harmonic oscillator. 
Therefore, we can write  $\phi(n, k , y) =\sum_{l =0}^\infty  \phi( n, k , l ) \psi_l(y)$  and we obtain the eigen spectrum of the
Laplacian on the plane wave to be given by 
\begin{eqnarray}
	\lambda &=& -\left( \frac{ 8\pi n i  r k }{  \beta}  +  4k^2  + \hat m^2 r^2  + 4k \big(l + \frac{1}{2} \big) \right) , 
	\\ \nonumber
	&& n \in \mathbb{Z},  \quad  l = 0, 1, 2, 3, \cdots , \quad  k \geq 0 .
\end{eqnarray}
We can now write the partition function of a massive scalar on the pp wave background as 
\begin{eqnarray}
	\log Z_{pp} =-  \sum_{l = 0}^\infty \sum_{n = -\infty}^\infty   \int_0^\infty \frac{dk}{2\pi} 
	\log\left[  4k  ( l + \frac{1}{2} )  + 4  k^2  +\hat m^2 r^2 + \frac{i 8\pi nr k }{\beta} \right].
\end{eqnarray}
Here we have also included the contribution from the complex conjugate  modes $\phi^*(n, k, l)$, this removes the  overall factor of 
$\frac{1}{2}$ in the partition function. 

To proceed with the analysis of the partition function, we first   scale out the factor of $4 k$ from the logarithm
using the zeta function regularization. 
We obtain 
\begin{eqnarray}
	\log Z_{pp} =- \sum_{l = 0}^\infty \sum_{n = -\infty}^\infty   \int_0^\infty \frac{dk}{2\pi} 
	\log\left({( l + \frac{1}{2} )} + k  +  \frac{\hat m^2r^2 }{4k}   +  i \frac{2\pi nr }{\beta}  \right) .
\end{eqnarray}
Now we organize the sum over Matsubara frequencies as, using $\mathbf{m}=\hat m/2$
\begin{eqnarray}\label{z0+z3}
	\log Z_{pp}  &=&- \frac{1}{2}\sum_{l = 0}^\infty \sum_{n = -\infty}^\infty   \int_0^\infty \frac{dk}{2\pi} 
	\log\left(  \Big[ {( l + \frac{1}{2} )}  + k  +  \frac{ \mathbf{m}^2r^2 }{k} \Big]^2    +  \Big[ \frac{2\pi nr }{\beta} \Big]^2   \right)  .
\end{eqnarray}
The sum over the Matsubara frequencies  can be done using the formula  in \cite{Klebanov:2011uf}  which results in 
\begin{align}\label{z1+z2}
	\log Z_{pp} &=-\frac12\sum_{l=0}^\infty \int_0^\infty \frac{dk}{\pi} \log \Big[2\sinh \Big\{\frac{\beta}{2r}\big(l+\frac{1}{2}+k+\frac{\mathbf m^2r^2}{k}\big)\Big\}\Big]\nonumber,\\
	&=-\frac12\sum_{l=0}^{\infty} \int_0^\infty \frac{dk}{\pi} \frac{\beta}{2r}\Big(l+\frac{1}{2}+k+\frac{\mathbf m^2r^2}{k}\Big)-\frac12\sum_{l=0}^{\infty} \int_0^\infty \frac{dk}{\pi}\log\Big[1-e^{-\frac{\beta}{r}(l+\frac{1}{2}+k+\frac{\mathbf m^2r^2}{k})}\Big]\nonumber,\\
	&\equiv- \frac12
	(\log Z_1+\log Z_2).
\end{align}

\subsubsection*{Evaluation of $\log Z_1$}

The first term in the above expression, denoted by $\log Z_1$, diverges and  therefore needs to be regulated. 
We regulate it by writing the sum over $l$ as a $\zeta$ function by introducing 
\begin{align}
	\log Z_1 (\alpha) &=\frac{\beta}{r} \sum_{l=0}^\infty \int_0^\infty\frac{dk}{2\pi}\Bigg[\frac{1}{\Gamma (-\alpha )}    \int_0^{\infty } t^{-\alpha -1} e^{-t \left(l+\frac{1}{2}+k+\frac{\mathbf m^2r^2}{k}\right)} \, dt\Bigg]_{\alpha=1}
	&\equiv \log Z_1(\alpha)|_{\alpha=1}.
\end{align}
The  $k$-integral and the sum over $l$ can be done and we obtain 
\begin{align}
	\log Z_1(\alpha)=\frac{\beta \mathbf m }{2 \pi  \Gamma (-\alpha )}\int_0^\infty dt\, t^{-\alpha -1} \text{csch}\left(\frac{t}{2}\right) K_1(2\mathbf m r t).
\end{align}
Using the following expansion for $1/\sinh(t/2)$
\begin{align}
	\frac{1}{\sinh(t/2)}=   \sum _{n=0}^{\infty } \frac{\left(2^{2-2 n} \left(1-2^{2 n-1}\right) B_{2 n}\right) }{(2 n)!}t^{2 n-1}.
\end{align}
One gets,
\begin{align}\label{Z1}
	\log Z_1(\alpha)&=\frac{\mathbf m\beta }{2\pi\Gamma (-\alpha )}  
	\sum _{n=0}^{\infty } \frac{\left(2^{2-2 n} \left(1-2^{2 n-1}\right) B_{2 n}\right) }{(2 n)!}
	\int_{0}^\infty t^{2n-\alpha -2}  K_1(2\mathbf m r t) dt.
\end{align}
We expand the Bessel  function at large argument  as given below \footnote{We could also perform the integration over $t$ directly, which 
	 leads to the same conclusion.}
\begin{align}
	K_1(2\mathbf mrt)=\sum_{k=0}^\infty \frac{(-1)^{k+1} 2^{-2 k-1} \Gamma \left(k-\frac{1}{2}\right) \Gamma \left(k+\frac{3}{2}\right) e^{-2 \mathbf mr t} }{\sqrt{\pi } k! (\mathbf m r t)^{k+\frac{1}{2}}}.
\end{align}
We substitute this expansion  in equation \eqref{Z1} to obtain
\begin{align}
	\log Z_1(\alpha)
	=	\sum_{n,k=0}^{\infty}
	\frac{\beta\mathbf m  (-1)^k  \left(4^n-2\right) B_{2 n} \Gamma \left(k-\frac{1}{2}\right) \Gamma \left(k+\frac{3}{2}\right)  \Gamma \left(2 n-k-\alpha -\frac{3}{2}\right) (\mathbf m r)^{\alpha -2 n+1}}{2^{4n+k-\alpha-\frac{1}{2}}\pi ^{3/2} k! (2 n)! \Gamma (-\alpha )}.
\end{align}
It is now easy to see that the above expression admits an expansion around $\alpha =1$, which does not have a pole. 
Expanding around  $\alpha=1$ and summing over $k$ from $0$ to $\infty$ we obtain
\begin{eqnarray}
	\lim_{\alpha\to 1}\log Z_1(\alpha)&=& (\alpha -1) \sum_{n=0}^{\infty}
	\frac{\sqrt{\pi }  \beta  \mathbf m^3 \left(4^n-2\right) r^2 B_{2 n} \, _2\tilde{F}_1\left(\frac{3}{2},4-2 n;\frac{7}{2}-2 n;-1\right) \sec (2 \pi  n)}{2^{4n-3}(\mathbf mr)^{2n}\Gamma (2 n+1)}, \nonumber \\ \nonumber
	& =&    (\alpha -1) \sum_{n=0}^{\infty}
	\frac{\sqrt{\pi }  \beta  \mathbf m^3 \left(4^n-2\right) r^2 B_{2 n} \;  _2{F}_1\left(\frac{3}{2},4-2 n;\frac{7}{2}-2 n;-1\right) }{2^{4n-3}(\mathbf mr)^{2n} ( 2n !) \Gamma\big( \frac{7}{2} -2 n \big) } \nonumber\\&&\qquad\qquad\qquad\qquad\qquad\qquad\qquad\qquad\qquad+ O( (\alpha -1)^2). 
\end{eqnarray}
Here $_2\tilde F_1(x)$ denotes the regularised hypergeometric function.
Therefore the regulated sum results in 
\begin{align}\label{z1 zero}
	\log Z_1=0.
\end{align}

\subsubsection*{Evaluation of $\log Z_2$}

We now evaluate the term $\log Z_2$ defined in
\eqref{z1+z2}.  For $\beta, l , \mathbf{ m}  r>0$, the expansion of the logarithm is
absolutely convergent, so the $l$-sum and the $k$-integral may be performed
term by term.  One obtains
\begin{equation}
	\log Z_2=-\sum_{l=0}^\infty\int_0^\infty\frac{\dd k}{\pi}
	\sum_{n=1}^\infty\frac1n
	\e^{-\frac{n\beta} r[l+\frac12+k+\frac{\mathbf m^2r^2}k]}.
	\label{eq:plane-wave-Z2-series}
\end{equation}
In \eqref{eq:plane-wave-Z2-series}, the $l$-sum gives
$[2\sinh(n\beta/2 r)]^{-1}$, while the $k$-integral gives the Bessel function. 
We obtain
\begin{align}\label{eq:plane-wave-Z2-summed}
	\log Z_2=-\frac{\mathbf m r }{\pi  }
	\sum_{n=1}^\infty
	\frac{	K_1(2 \mathbf m n \beta )}{ n\sinh\frac{\beta  n}{2 r}}.
\end{align}
Combining \eqref{z1 zero} and
\eqref{eq:plane-wave-Z2-summed} 
according to the decompositions \eqref{z0+z3} and \eqref{z1+z2} gives
\begin{equation}
	\log Z(\mathbf m;\beta, r)_{pp}
	=
	\frac{\mathbf m r }{2\pi  }
	\sum_{n=1}^\infty
	\frac{	K_1(2 \mathbf m n \beta )}{ n\sinh\frac{\beta  n}{2 r}}.
	\label{eq:plane-wave-logZ-final}
\end{equation}
As a check of this result,  observe that 
\begin{eqnarray}
	\lim_{ \mathbf m \rightarrow 0}  \log Z(\mathbf m;\beta, r)_{pp} = \frac{r}{4\pi \beta}\sum_{n=1}^\infty \frac{1}{n^2 \sinh\frac{\beta  n}{2 r}}.
\end{eqnarray}
Multiplying this with the volume of the pp wave $\frac{2\pi }{\epsilon}$ as in \cite{Komargodski:2026ain}, the partition function precisely agrees with 
equation (4.9) of \cite{Komargodski:2026ain}, as well as with the result of our independent calculation shown in the leading term of \eqref{eq:free-small-x-full}, with the identification $x^2=2\epsilon$.

We are now ready to find the fixed point of the large $N$,  vector model  on the pp wave. 
The saddle point is given by  minimizing the partition function 
with respect to $\mathbf m$.  
From 
\eqref{eq:plane-wave-logZ-final} we obtain the gap equation
\begin{equation}
	\frac{\partial\log Z_{pp} }{\partial \mathbf m}
	=-\mathbf m\beta    r\sum_{n=1}^\infty\frac{  K_0(2 \mathbf m n \beta ) }{\pi \sinh\frac{\beta n}{2r}}=0.
	\label{eq:plane-wave-gap}
\end{equation}
The only real and non-negative solution  for $\mathbf m$ satisfying the above equation for $\beta, r >0$ is 
$\mathbf m=0$.  
This implies that the free energy of the large $N$ critical  vector model  on $S^1\times S^2$ 
at $x\rightarrow 0$  or $\mu^2 r^2 + 1\rightarrow 0$
coincides with that of the massless theory. 
This provides a consistency check of the explicit  and detailed calculation of the behaviour   of the $\frac{r^2}{\beta^2}$  in the section \ref{sec:interacting}. 

Our calculation of the partition function of the large $N$ critical model on the pp-wave  geometry also predicts  that 
as $x$ is dialled to zero, all sub-leading coefficients  in the $\frac{\beta}{r}$ expansion  of the partition function of the critical 
vector model at large $N$ also develop poles at $x=0$ and their residue coincides with that of the massless theory. 
It will be interesting to verify this prediction by similar calculations as done for the $r^2/\beta^2$ coefficient in this paper.

\section{Discussions}
\label{sec:conclusions}
In this work we have examined the free energy of a large $N$ critical vector model of scalars with angular potential in $(2+1)\, d$. The model is restricted so  that the large $N$ dynamics is completely described by the constant mode of the auxiliary field of Hubbard-Stratonovich transformation, even at non-zero angular potential. We have computed the leading high temperature contribution to free energy as systematic  expansions at  vanishing angular potential $\hat\mu r=0$ and as well as at the limit $\hat\mu^2 r^2 =1$. We also have shown that the answers at these two different limits interpolate via a smooth curve obtained numerically in the range $\hat\mu r\in (0,1)$.
The free energy at $\hat\mu r=0$ reproduces the known result in absence of the angular potential \cite{Sachdev:1993pr}. 
On the other hand it exhibits a simple pole at $\hat\mu^2r^2=1$ as its  leading contribution, thus verifying the predictions from the thermal effective field theory \cite{Anand:2025mfh}. 
The pole contribution at  the non-trivial fixed point coincides with that of the free theory, while the sub-leading terms in the series about $\hat\mu^2r^2=1$ are non-analytic, distinguishing it from the free theory result.  This non-analytic behaviour originates from the non-analyticity of the thermal mass satisfying the gap equation as a series expansion about $\hat\mu^2r^2=1$.  
It would be interesting to study how to incorporate such terms within thermal effective field theory.  
Such logarithmic terms were also present for the free massive theory
in \cite{Anand:2025mfh}.
Our results can provide an explicit example to test such a generalisation of  the thermal effective field theory.

As we have emphasized we have studied  the vector  model coupled  only to the constant mode of the Hubbard-Stratonovich 
auxiliary field.  Integrating out this field results in a non-local quartic interaction of the original $O(N)$ model. 
In spite of this we find the predictions based on  local  effective theory of \cite{Anand:2025mfh} is valid. 
This is a bit surprising, and worth investigating further in other models.  The interpolation of the free energy of the standard 
$O(N)$ model at large $N$ at its non-trivial IR  fixed point and zero angular  potential  to  the free energy of the 
free theory or the UV theory at $\hat \mu^2 r^2 =1$  via  the model presented here is interesting and may help to understand the 
behaviour of the standard $O(N)$ model with angular velocity.

We also have applied the proposal of \cite{Komargodski:2026ain} to read out the residue at the pole at $\hat\mu^2r^2=1$
by studying this vector model  on a pp-wave geometry.  The result for the residue agrees with our 
direct computation. 
The methods developed in this paper for evaluating partition functions  at non-zero angular 
 potential  and solving the gap equations can be generalised for other vector models involving more general interactions and  higher dimensions \cite{Romatschke:2019ybu,Petkou:2021zhg,Grable:2022swa,Kumar:2025txh} and also PT symmetric theories \cite{Lawrence:2023woz,Grable:2023paf}.

In this paper we  have focused on the leading high temperature behaviour of the free energy in presence of the angular potential as a direct computation. It would be interesting to observe similar poles in the sub-leading orders  
in the free energy as a systematic high temperature expansion as developed for the case without angular velocity in 
\cite{David:2024pir}.  Our analysis of this model in the pp-wave geometry  predicts that these poles must coincide with that 
of  free theory. 
The generalization of the analysis   for the squashed sphere \cite{Parmentier:2026aqh} is an  interesting direction for further investigation. 
In this paper  we have dealt with the large $N$, critical vector  model without singlet constraint. The model with the singlet constraint is of particular interest as it undergoes Gross-Witten-Wadia transition \cite{Shenker:2011zf}. 
It would be interesting to  study the behaviour of this  phase transition in presence of the angular potential.

Finally it would be interesting to develop the methods of thermal bootstrap to obtain the behaviour of one point functions of  primaries 
of conformal field theories in general and the $O(N)$ model in particular  at finite angular potential.

\section*{Acknowledgments}
The use of Feynman parametrization in \eqref{eq:branch-feynman-parametrized} was suggested by ChatGPT and proved to be instrumental in the further analysis.
 ChatGPT and Codex were also used to assist with the subsequent algebraic manipulations in Section \ref{sec:interacting}, parts of consistency checks, numerical analysis and  manuscript organization. 
 All such steps were carried out under human supervision and were independently verified at every stage, either by  using \textit{Mathematica} or independent calculations.

 We thank Shiraz Minwalla,  Vinayak Mishra and Jyotirmoy Mukherjee 
  for pointing out the non-uniformity of the thermal mass of the standard $O(N)$ model  at large $N$ 
 with angular potential  during discussions and correspondences.  This led to the clarification of the model addressed in  the  paper. 
JRD is partially supported by ANRF, India: grant no. ANRF/ARGM/2025/00054/TS. SK is supported by the National Natural Science Foundation of China (NSFC) under Grants No. 12247103.

\appendix

\section{High-temperature expansion  of the free-theory pole}
\label{app:free-highT}
In this appendix we reorganize the residue at the pole at $x=0$ for the free energy in free theory given in \eqref{eq:free-small-x-full} as a small $\beta/r$ expansion to obtain the result \eqref{eq:free-summed-highT-few}.
Let us consider the coefficient of the $x^{-2}$ pole in \eqref{eq:free-small-x-full} as given below
\begin{equation}
	C_{\rm free}\!\left(\frac{\beta}{r}\right)
	\equiv
	\lim_{x\to0}x^2\log Z_{\rm free}
	=
	\frac{r}{\beta}
	\sum_{n=1}^{\infty}
	\frac{1}{n^2\sinh\!\left(n\beta/2r\right)}.
	\label{eq:app-free-def}
\end{equation}
Now using the following geometric series 
\begin{equation}
	\frac{1}{\sinh\!\left(n\beta/2r\right)}
	=
	2\sum_{j=0}^{\infty}
	\e^{-(j+\frac12)n\beta/r},
\end{equation}
together with the Mellin representation
\begin{equation}
	\e^{-t}
	=
	\frac{1}{2\pi i}
	\int_{c-i\infty}^{c+i\infty}\dd s\,
	\Gamma(s)t^{-s},
	\qquad c>0,
\end{equation}
the sums over $n$ and $j$ can be performed with the use of the standard formulae
\begin{equation}
	\sum_{n=1}^{\infty}\frac{1}{n^{s+2}}
	=
	\zeta(s+2),
	\qquad
	\sum_{j=0}^{\infty}
	\frac{1}{(j+\frac12)^s}
	=
	(2^s-1)\zeta(s),
\end{equation}
one obtains
\begin{equation}
	C_{\rm free}\!\left(\frac{\beta}{r}\right)
	=
	\frac{1}{2\pi i}
	\int_{c-i\infty}^{c+i\infty}\dd s\,
	2(2^s-1)\Gamma(s)\zeta(s)\zeta(s+2)
	\left(\frac{\beta}{r}\right)^{-s-1},
	\qquad c>1.
	\label{eq:app-free-MB}
\end{equation}

The high-temperature expansion follows by shifting the contour in
\eqref{eq:app-free-MB} to the left.  The pole at $s=1$ gives the leading
term $2\zeta(3)(r/\beta)^2$, while the double pole at $s=-1$ produces
the logarithmic and constant contributions.  The remaining poles at
$s=-2k-1$, $k\geq1$, give
\begin{equation}
	c_k
	=
	\frac{
		2\bigl(1-2^{-2k-1}\bigr)
		\zeta(-2k-1)\zeta(1-2k)}
	{(2k+1)!}.
	\label{eq:app-free-ck}
\end{equation}
Therefore
\begin{align}
	C_{\rm free}\!\left(\frac{\beta}{r}\right)
	\sim{}&
	2\zeta(3)\frac{r^2}{\beta^2}
	+\frac{1}{12}\log\!\frac{\beta}{r}
	+\zeta'(-1)
	+\frac{\log2-1}{12}
+	\sum_{k=1}^{\infty}
	c_k\left(\frac{\beta}{r}\right)^{2k}.
	\label{eq:app-free-all-orders}
\end{align}
Finally, the functional equation of the Riemann zeta function yields
\begin{equation}
	c_k
	\sim
	-\frac{8(2k-1)!}{(2\pi)^{4k+2}},
	\qquad k\to\infty.
	\label{eq:app-free-large-order}
\end{equation}
The factorial growth of the coefficients shows that
\eqref{eq:app-free-all-orders} is an asymptotic expansion in
$\beta/r$ with zero radius of convergence.
Thus the calculation above reproduces the high-temperature
pole expansion stated in the equation
\eqref{eq:free-summed-highT-few}.

\section{Small $\tilde m \beta x$ expansion of the gap equation}
\label{app:bessel-poisson}

In this appendix we show the derivation of the equation \eqref{eq:I-all-orders} from \eqref{eq:BP-bessel-sum}.
We had the equation \eqref{eq:BP-bessel-sum} as
\begin{equation}
	-I_\infty(\tilde m\beta x)
	=\sum_{n=1}^\infty\frac{K_0(n\tilde m\beta x)}{n}.
	\label{eq:BP-bessel-sum-appen}
\end{equation}
First we apply the following derivative operation to the above equation, this will allow us to rearrange the resulting series with the use of Poisson resummation, 
\begin{equation}
	\left[(\tilde m\beta x)\frac{\partial}{\partial(\tilde m\beta x)}\right]^2
	[-I_\infty(\tilde m\beta x)]
	=(\tilde m\beta x)^2\sum_{n=1}^\infty nK_0(n\tilde m\beta x).
	\label{eq:BP-differential-reduction}
\end{equation}
and at the very end we can recover the reorganized series expansion for  $  -I_\infty (\tilde m \beta x)$ by integrating it appropriately.\\

Noting the fact that the summand in \eqref{eq:BP-bessel-sum-appen} is even in $n$, we can apply 
the Poisson summation formula, as given below 
\begin{equation}
	\sum_{n=-\infty}^{\infty}f(n)
	=
	\sum_{p=-\infty}^{\infty}\widehat f(2\pi p),
	\label{eq:BP-Poisson-general}
	\qquad {\rm with}\qquad \widehat f(b)=\int_{-\infty}^{\infty}\dd y\,\e^{-iby}f(y),
\end{equation}
to obtain the following equality
\begin{align}
	2\sum_{n=1}^{\infty}nK_0(n\tilde m\beta x)
	={}&
	\frac{2}{(\tilde m\beta x)^2}
	+
	4\sum_{p=1}^{\infty}
	\Bigg[
	\frac{1}{(\tilde m\beta x)^2+(2\pi p)^2}
	-
	\frac{2\pi p\,
		\operatorname{arcsinh}[2\pi p/(\tilde m\beta x)]}
	{[(\tilde m\beta x)^2+(2\pi p)^2]^{3/2}}
	\Bigg].
	\label{eq:BP-Poisson-sum}
\end{align}
Now we will expand the summand in small $\tilde m \beta x$. We need to combine the following expansions to do so
\begin{align}
&	\operatorname{arcsinh}\!\frac{2\pi p}{\tilde m\beta x}
	={\log\!\frac{4\pi p}{\tilde m\beta x}}
	+\sum_{j=1}^\infty
	\frac{(-1)^{j+1}}{2j\,4^j}\binom{2j}{j}
	\left(\frac{\tilde m\beta x}{2\pi p}\right)^{2j},
	\nonumber\\
&	\left[1+\left(\frac{\tilde m\beta x}{2\pi p}\right)^2\right]^{-1}
	=\sum_{j=0}^\infty(-1)^j
	\left(\frac{\tilde m\beta x}{2\pi p}\right)^{2j},
	\nonumber\\
&	\left[1+\left(\frac{\tilde m\beta x}{2\pi p}\right)^2\right]^{-3/2}
	=\sum_{j=0}^\infty
	\frac{(-1)^j(2j+1)!}{4^j(j!)^2}
	\left(\frac{\tilde m\beta x}{2\pi p}\right)^{2j}.
	\label{eq:BP-denom32-expansion}
\end{align}
The summand in
\eqref{eq:BP-Poisson-sum} can be expressed as the following series with use of the above expansion formulae, followed by a reorganization of the summation indices
\begin{align}
	&\frac{1}{(\tilde m\beta x)^2+(2\pi p)^2}
	-\frac{2\pi p\,
		\operatorname{arsinh}[2\pi p/(\tilde m\beta x)]}
	{[(\tilde m\beta x)^2+(2\pi p)^2]^{3/2}}
	\nonumber\\
	&\quad=
	\frac{1}{(2\pi p)^2}
	\left[
	1-\log\!\frac{4\pi p}{\tilde m\beta x}
	\right]
	+
	\frac{1}{(2\pi p)^2}
	\sum_{m=1}^{\infty}
	\Bigg[
	(-1)^m
	-\frac{(-1)^m(2m+1)!}{4^m(m!)^2}
	\log\!\frac{4\pi p}{\tilde m\beta x}
	\nonumber\\
	&\hspace{40mm}
	-\sum_{j=1}^{m}
	\frac{1}{2j\,4^j}
	\binom{2j}{j}
	\frac{(-1)^{m+1}(2m-2j+1)!}
	{4^{m-j}[(m-j)!]^2}
	\Bigg]
	\left(
	\frac{\tilde m\beta x}{2\pi p}
	\right)^{2m}.
	\label{eq:BP-Cauchy-coefficient}
\end{align}
For $m\geq1$, the coefficient of
$(\tilde m\beta x/2\pi p)^{2m}$ in
\eqref{eq:BP-Cauchy-coefficient} is therefore
\begin{align}
	\operatorname{Coeff}_{m}
	={}&
	(-1)^m
	\Bigg[
	1+
	\sum_{j=1}^{m}
	\frac{1}{2j\,4^j}
	\binom{2j}{j}
	\frac{(2m-2j+1)!}
	{4^{m-j}[(m-j)!]^2}
	-\frac{(2m+1)!}{4^m(m!)^2}
	\log\!\frac{4\pi p}{\tilde m\beta x}
	\Bigg].
	\label{eq:BP-m-coefficient-rearranged}
\end{align}
The logarithmic contribution already carries the factorial factor
$(2m+1)!/[4^m(m!)^2]$.  We therefore normalize the remaining
nonlogarithmic part by the same factor, and call it
\begin{equation}
	R_m
	\equiv
	\frac{4^m(m!)^2}{(2m+1)!}
	\left[
	1+
	\sum_{j=1}^{m}
	\frac{1}{2j\,4^j}
	\binom{2j}{j}
	\frac{(2m-2j+1)!}
	{4^{m-j}[(m-j)!]^2}
	\right].
	\label{eq:BP-Rm-def}
\end{equation}
Equation \eqref{eq:BP-m-coefficient-rearranged} then becomes
\begin{equation}
	\operatorname{Coeff}_{m}
	=
	(-1)^m
	\frac{(2m+1)!}{4^m(m!)^2}
	\left[
	R_m-\log\!\frac{4\pi p}{\tilde m\beta x}
	\right].
	\label{eq:BP-m-coefficient-Rm}
\end{equation}
A direct subtraction of consecutive coefficients gives
\begin{align}
	R_m-R_{m-1}
	=
	\frac{1}{2m(2m+1)}
+
	\frac{m!(m-1)!}{2(2m+1)!}
	\left[
	\sum_{j=1}^{m-1}
	\binom{2j}{j}
	\binom{2m-2j}{m-j}
	-4^m
	\right].
	\label{eq:BP-Rm-difference}
\end{align}
Now we have the following identity
\begin{equation}
	\sum_{j=0}^{m}
	\binom{2j}{j}
	\binom{2m-2j}{m-j}
	=
	4^m,
	\label{eq:BP-Vandermonde}
\end{equation}
and therefore
\begin{equation}
	\sum_{j=1}^{m-1}
	\binom{2j}{j}
	\binom{2m-2j}{m-j}
	=
	4^m-2\binom{2m}{m}.
	\label{eq:BP-Vandermonde-reduced}
\end{equation}
Substitution into \eqref{eq:BP-Rm-difference} gives
\begin{equation}
	R_m-R_{m-1}
	=
	-\frac{1}{2m(2m+1)},
	\qquad
	R_0=1.
	\label{eq:BP-Rm-recurrence}
\end{equation}
Hence
\begin{equation}
	R_m
	=
	1-\sum_{q=1}^{m}\frac{1}{2q(2q+1)}
	=
	H_{2m+1}-H_m,
	\qquad
	H_n\equiv\sum_{q=1}^{n}\frac1q .
	\label{eq:BP-harmonic-reduction}
\end{equation}
The coefficient \eqref{eq:BP-m-coefficient-Rm} therefore reduces to
\begin{equation}
	\operatorname{Coeff}_{m}
	=
	-\frac{(-1)^m(2m+1)!}{4^m(m!)^2}
	\left[
	\log\!\frac{4\pi p}{\tilde m\beta x}
	+H_m-H_{2m+1}
	\right],
	\qquad m\geq1.
	\label{eq:BP-m-coefficient-final}
\end{equation}
Substituting this result into \eqref{eq:BP-Cauchy-coefficient} gives
\begin{align}
	&\frac{1}{(\tilde m\beta x)^2+(2\pi p)^2}
	-\frac{2\pi p\,
		\operatorname{arsinh}[2\pi p/(\tilde m\beta x)]}
	{[(\tilde m\beta x)^2+(2\pi p)^2]^{3/2}}
=
	\frac{1-\log[4\pi p/(\tilde m\beta x)]}{(2\pi p)^2}
	\nonumber\\
	&\qquad\qquad
	-\frac{1}{(2\pi p)^2}
	\sum_{m=1}^{\infty}
	\frac{(-1)^m(2m+1)!}{4^m(m!)^2}
	\left[
	\log\!\frac{4\pi p}{\tilde m\beta x}
	+H_m-H_{2m+1}
	\right]
	\left(
	\frac{\tilde m\beta x}{2\pi p}
	\right)^{2m}.
	\label{eq:BP-summand-m}
\end{align}
The $p$-sum in \eqref{eq:BP-Poisson-sum} can now be performed directly.  Using
\begin{align}
&	\sum_{p=1}^{\infty}
	\frac{
		\log[4\pi p/(\tilde m\beta x)]
		+H_{k-1}-H_{2k-1}}
	{(2\pi p)^{2k}}\nonumber\\
&\qquad\qquad	=
	\frac{(-1)^{k+1}B_{2k}}{2(2k)!}
	\left[
	\log\!\frac{2}{\tilde m\beta x}
	+\log(2\pi)
	+H_{k-1}-H_{2k-1}
	-\frac{\zeta'(2k)}{\zeta(2k)}
	\right],
	\label{eq:BP-p-sum}
\end{align}
where
\begin{equation}
	\sum_{p=1}^{\infty}\frac{1}{(2\pi p)^{2k}}
	=
	\frac{(-1)^{k+1}B_{2k}}{2(2k)!},
	\qquad
	\sum_{p=1}^{\infty}\frac{\log p}{p^{2k}}
	=
	-\zeta'(2k),
	\qquad k\geq1,
\end{equation}
Thus the Poisson-resummed series becomes
\begin{align}
	&\left[
	(\tilde m\beta x)
	\frac{\partial}{\partial(\tilde m\beta x)}
	\right]^2
	[-I_\infty(\tilde m\beta x)]
	\nonumber\\
	&\quad=
	1-\sum_{k=1}^{\infty}
	\frac{2kB_{2k}(\tilde m\beta x)^{2k}}
	{4^k(k!)^2}
	\left[
	\log\!\frac{2}{\tilde m\beta x}
	+\log(2\pi)
	+H_{k-1}-H_{2k-1}
	-\frac{\zeta'(2k)}{\zeta(2k)}
	\right].
	\label{eq:BP-differential-series}
\end{align}
It remains to integrate the above expression appropriately to recover $-I_\infty(\tilde m \beta x)$. We directly integrate the above expression which gives rise to two constants of integration to be determined, the result of the integration is given below
\begin{align}
	-I_\infty(\tilde m\beta x)
	={}&
	\frac12\log^2\!\frac{2}{\tilde m\beta x}
	+C_1+C_2\log\!\frac{2}{\tilde m\beta x}
	\nonumber\\
	&-
	\sum_{k=1}^{\infty}
	\frac{B_{2k}(\tilde m\beta x)^{2k}}
	{2k\,4^k(k!)^2}
	\left[
	\log\!\frac{2}{\tilde m\beta x}
	+\log(2\pi)
	+H_k-H_{2k-1}
	-\frac{\zeta'(2k)}{\zeta(2k)}
	\right].
	\label{eq:BP-before-constants}
\end{align}
$C_1, C_2$ are the constants of integration to be determined. Here we have used that fact that
\begin{equation}
	H_{k-1}+\frac1k=H_k,
\end{equation}
In the remainder of this appendix we will fix these integration constant $C_1$ and $C_2$ by analyzing the original form of the integral $I_{\infty}(\tilde{m}\beta x)$. Let us consider the original expression for the integral $I_{\infty}(\tilde m \beta x)$ from \eqref{eq:I-finite}, with a change in the integration variable $u=\tilde{m}\beta z$ 
\begin{equation}
	-I_\infty(\tilde m\beta x)
	=
	-\int_{\tilde m\beta x}^{\infty}\dd u\,
	\frac{\log(1-\e^{-u})}
	{\sqrt{u^2-(\tilde m\beta x)^2}}.
	\label{eq:BP-original-integral}
\end{equation}
We split the integration range at $u=1$ and isolate the singular part
of the numerator on $0<u<1$ by the exact identity
\begin{equation}
	\log(1-\e^{-u})
	=
	\log u
	+
	\log\!\left(\frac{1-\e^{-u}}{u}\right).
	\label{eq:BP-log-split}
\end{equation}
This gives
\begin{align}
	-I_\infty(\tilde m\beta x)
	={}&
	-\int_{\tilde m\beta x}^{1}\dd u\,
	\frac{\log u}{\sqrt{u^2-(\tilde m\beta x)^2}}
	-
	\int_{\tilde m\beta x}^{1}\dd u\,
	\frac{\log[(1-\e^{-u})/u]}
	{\sqrt{u^2-(\tilde m\beta x)^2}}
	\nonumber\\
	&\qquad\qquad\qquad\qquad-
	\int_{1}^{\infty}\dd u\,
	\frac{\log(1-\e^{-u})}
	{\sqrt{u^2-(\tilde m\beta x)^2}}.
	\label{eq:BP-C0-split}
\end{align}
The first term contains the complete logarithmic singularity, while
the rest of the terms are regular at $\tilde m \beta x=0$ and contributes $O(1)$ constants as  leading contributions. One can compute the integral in the first term exactly, thereby take the limit $\tilde m \beta x\to0  $ to the resulting expression as shown below
\begin{align}
&	-\int_{\tilde m\beta x}^{1}\dd u\,
	\frac{\log u}{\sqrt{u^2-(\tilde m\beta x)^2}}\nonumber
\\&	=
	\frac12\log^2\!\frac{2}{\tilde m\beta x}
	-\frac14
	\log^2\!\left(
	\frac{1+\sqrt{1-\tilde m^2\beta^2x^2}}{2}
	\right)
	+\frac12\Li_2\!\left(
	\frac{1-\sqrt{1-\tilde m^2\beta^2x^2}}{2}
	\right)
	-\frac{\pi^2}{24},\nonumber\\
	&=\frac{1}{2}\log^2(\frac2{\tilde{m}\beta x})-\frac{\pi^2}{24}+O((\tilde m\beta x)^2)
	\label{eq:BP-log-singular}
\end{align}
Now using this and the fact that the 2nd and the last term from \eqref{eq:BP-C0-split} remains finite at $\tilde m\beta x\to0$ limit, we evaluate non-vanishing contribution of $-I_\infty(\tilde{m}\beta x)$ at $\tilde m\beta x\to0$, as the following
\begin{equation}
\lim_{\tilde m \beta x\to 0}[	-I_\infty(\tilde m\beta x)]
	=
	\frac12\log^2\!\frac{2}{\tilde m\beta x}
	+C_0,
	\label{eq:BP-small-argument-limit}
\end{equation}
Here we have suppressed all the terms those tends to zero at this limit, with $C_0$ being the $O(1)$ constant term, obtained by combining the constant from \eqref{eq:BP-log-singular} and the $\tilde m \beta x\to 0$ limit of the integrals in the 2nd and the last term in \eqref{eq:BP-C0-split}, as given below
 \begin{equation}
	C_0
	=
	-\frac{\pi^2}{24}
	-\int_0^1\dd u\,
	\frac{\log[(1-\e^{-u})/u]}{u}
	-\int_1^\infty\dd u\,
	\frac{\log(1-\e^{-u})}{u}.
	\label{eq:BP-C0}
\end{equation}
The integrals in the above equation are convergent and will be evaluated later. Thus, comparing the equation \eqref{eq:BP-before-constants} and \eqref{eq:BP-small-argument-limit}, we identify the constants $C_1$ and $C_2$ as the following
\begin{equation}
	C_1=C_0,
	\qquad
	C_2=0.
\end{equation}
The constant $C_0$ in \eqref{eq:BP-C0} can be evaluated as a compact analytic answer following the steps shown below. Let us use the following standard integration result
\begin{equation}
	\int_0^\infty \dd u\,u^{s-1}\log(1-\e^{-u})
	=
	-\Gamma(s)\zeta(1+s),
	\qquad {\rm Re}\ s>0.
\end{equation}
Now since we have
\begin{equation}
	\int_0^1\dd u\,u^{s-1}\log u
	=
	-\frac{1}{s^2},
\end{equation}
the convergent combination appearing in $C_0$ can be obtained as
\begin{align}
	&\int_0^1\dd u\,
	\frac{\log[(1-\e^{-u})/u]}{u}
	+
	\int_1^\infty\dd u\,
	\frac{\log(1-\e^{-u})}{u}
=
	\lim_{s\to0}
	\left[
	-\Gamma(s)\zeta(1+s)+\frac{1}{s^2}
	\right].
	\label{eq:C0-reg-combination}
\end{align}
Using the known expansions for Gamma and Zeta function as given below
\begin{equation}
	\Gamma(s)
	=
	\frac1s-\gamma_{\mathrm E}
	+\left(
	\frac{\gamma_{\mathrm E}^2}{2}
	+\frac{\pi^2}{12}
	\right)s+O(s^2),
	\qquad
	\zeta(1+s)
	=
	\frac1s+\gamma_{\mathrm E}
	-\gamma_1 s+O(s^2),
\end{equation}
with $\gamma_{\mathrm E}$ is the Euler--Mascheroni constant, and
$\gamma_1$ is the first Stieltjes constant.
It then follows that
\begin{equation}\label{GammaZeta}
	-\Gamma(s)\zeta(1+s)
	=
	-\frac1{s^2}
	+\gamma_1
	+\frac{\gamma_{\mathrm E}^2}{2}
	-\frac{\pi^2}{12}
	+O(s).
\end{equation}
Therefore combining \eqref{GammaZeta} and \eqref{eq:C0-reg-combination} we have
\begin{equation}
	\int_0^1\dd u\,
	\frac{\log[(1-\e^{-u})/u]}{u}
	+
	\int_1^\infty\dd u\,
	\frac{\log(1-\e^{-u})}{u}
	=
	\gamma_1+\frac{\gamma_{\mathrm E}^2}{2}
	-\frac{\pi^2}{12},
\end{equation}
Hence substituting this in \eqref{eq:BP-C0} we obtain
\begin{equation}
	C_0
	=
	\frac{\pi^2}{24}
	-\frac{\gamma_{\mathrm E}^2}{2}
	-\gamma_1.
	\label{eq:BP-C0-closed}
\end{equation}
\section{Small {$x$} expansion of the thermal mass}
\label{app:thermal-mass-orders}

In this appendix, we solve the gap equation \eqref{eq:gap-expanded} as a small $x$ expansion with higher order terms
till the order of $x^4$.  For compactness we use the following notations
\begin{equation}
	M\equiv\tilde m(x)\beta,
	\qquad
	\Lambda\equiv\log\!\frac1x ,
	\label{eq:app-mass-definitions}
\end{equation}
In the following discussions it is understood that $M$ is a function of $x$ and $\beta$ and will not be displayed explicitly in order to streamline the presentation. Now we have, as $x\to0$,
\begin{equation}
	M\sim\Lambda^2,
	\qquad
	Mx\sim x\Lambda^2\longrightarrow0.
\end{equation}
With the use of  definitions in \eqref{eq:app-mass-definitions}, we can write down the gap equation \eqref{eq:gap-expanded}
 truncating
 after the $k=2$ term gives
\begin{align}
	M={}&\frac1{\sqrt{1-x^2}}
	\Bigg[
	\log^2\!\frac{2}{Mx}+2C_0
	-2(Mx)^2
	\left[
	\frac1{48}\log\!\frac{2}{Mx}+C_2
	\right]
	\nonumber\\
	&\hspace{2mm}
	-2(Mx)^4
	\left[
	-\frac1{7680}\log\!\frac{2}{Mx}+C_4
	\right]
	+O\!\left(
	(Mx)^6\log\!\frac{2}{Mx}
	\right)
	\Bigg]
+O\!\left(
	\frac{\e^{-M}}{M\sqrt{1-x^2}}
	\right).
	\label{eq:app-mass-gap-k2}
\end{align}
The logarithmic part of the $k$-th term scales as
$x^{2k}\Lambda^{4k+1}$; hence the first omitted algebraic sector is
$O(x^6\Lambda^{13})$.  We therefore can use the following ansatz for $M$ to solve the gap equation given in the above equation
\begin{equation}
	M
	=
	M_{\rm L}(x)
	+x^2A_2(x)
	+x^4A_4(x)
	+O\!\left(x^6\Lambda^{13}\right).
	\label{eq:app-mass-sector-ansatz}
\end{equation}
Here $M_{\rm L}(x)$ collects the terms with no explicit positive power
of $x$.  The coefficient functions $A_2(x)$ and $A_4(x)$ retain the
logarithmic dependence associated with the $x^2$ and $x^4$ sectors and
are therefore not constant Taylor coefficients.  We also define
\begin{equation}
	\ell(x)\equiv
	\log\!\frac{2}{M_{\rm L}(x)x}.
\end{equation}
The ansatz \eqref{eq:app-mass-sector-ansatz} gives the following
expansion of the mass-dependent logarithm and the geometric prefactor is also expanded at small $x$ in the following
\begin{align}
	\log\!\frac{2}{Mx}
	={}&
	\ell
	-x^2\frac{A_2}{M_{\rm L}}
	+x^4
	\left(
	\frac{A_2^2}{2M_{\rm L}^2}
	-\frac{A_4}{M_{\rm L}}
	\right)
	+O(x^6\Lambda^{11}),
	\nonumber\\
	\frac1{\sqrt{1-x^2}}
	={}&
	1+\frac{x^2}{2}
	+\frac{3x^4}{8}
	+O(x^6),
	\label{eq:app-mass-expansions}
\end{align}
Here and throughout the subsequent discussion, the $x$-dependence of $M_{\rm L}$, $A_2$, $A_4$, and $\ell$ is left implicit for notational simplicity.
Substitution of \eqref{eq:app-mass-sector-ansatz} and
\eqref{eq:app-mass-expansions} into \eqref{eq:app-mass-gap-k2}, followed
by matching explicit powers of $x$, gives
\begin{align}
	M_{\rm L}+x^2A_2+x^4A_4
	={}&\ell^2+2C_0
	+x^2\Bigg[
	\frac12\left(\ell^2+2C_0\right)
	-\frac{2\ell A_2}{M_{\rm L}}
	-M_{\rm L}^2\left(\frac{\ell}{24}+2C_2\right)
	\Bigg]
	\nonumber\\
	&+x^4\Bigg[
	\frac38\left(\ell^2+2C_0\right)
	-\frac{\ell A_2}{M_{\rm L}}
	-\frac{M_{\rm L}^2}{2}
	\left(\frac{\ell}{24}+2C_2\right)
	\nonumber\\
	&\hspace{17mm}
	+\frac{1+\ell}{M_{\rm L}^2}A_2^2
	-\frac{2\ell A_4}{M_{\rm L}}
	+M_{\rm L}A_2
	\left(\frac1{24}-\frac{\ell}{12}-4C_2\right)
	\nonumber\\
	&\hspace{17mm}
	+M_{\rm L}^4
	\left(\frac{\ell}{3840}-2C_4\right)
	\Bigg]
	+O\!\left(x^6\Lambda^{13}\right)
	+O\!\left(\frac{\e^{-M_{\rm L}}}{M_{\rm L}}\right).
	\label{eq:app-mass-gap-collected}
\end{align}
The final remainder in \eqref{eq:app-mass-gap-collected} is
exponentially smaller than every algebraic sector and does not affect
the order-by-order matching.\\
At order $x^0$, \eqref{eq:app-mass-gap-collected} reduces to
\begin{equation}
	M_{\rm L}
	=
	\ell^2+2C_0,
	\qquad
	\ell
	=
	\log\!\frac{2}{M_{\rm L}x}.
	\label{eq:app-mass-leading-sector}
\end{equation}
Ignoring the $C_0$ contribution in
\eqref{eq:app-mass-leading-sector} isolates the ProductLog
approximation, as given below
\begin{equation*}
	M_{\rm prodlog}(x)
=	
	4W_0^2\!\left(\frac1{\sqrt{2x}}\right),
\end{equation*}
which is the solution quoted in \eqref{Prodlog} of the main text.
Retaining $C_0$ in \eqref{eq:app-mass-leading-sector} defines the
improved leading sector $M_{\rm L}$ before any power-suppressed
corrections are included.  Its physical large-$M_{\rm L}$ branch has
the asymptotic behavior
\begin{equation}
	M_{\rm L}
	=
	\Lambda^2
	\left[
	1+O\!\left(
	\frac{\log\Lambda}{\Lambda}
	\right)
	\right],
	\qquad
	\ell
	=
	\Lambda
	\left[
	1+O\!\left(
	\frac{\log\Lambda}{\Lambda}
	\right)
	\right].
	\label{eq:app-mass-leading-orders}
\end{equation}

At order $x^2$, the expansion of the leading logarithm, the geometric
prefactor, and the direct $k=1$ term, all of these contribute.  Solving the
resulting equation for $A_2$ gives
\begin{equation}
	A_2(x)
	=
	\frac{\displaystyle
		\frac{M_{\rm L}}2
		-M_{\rm L}^2
		\left(
		\frac{\ell}{24}+2C_2
		\right)}
	{\displaystyle
		1+\frac{2\ell}{M_{\rm L}}}.
	\label{eq:app-mass-A2-exact}
\end{equation}
Together with \eqref{eq:app-mass-leading-orders}, this gives
\begin{equation}
	A_2(x)
	=
	-\frac{\Lambda^5}{24}
	\left[
	1+O\!\left(
	\frac{\log\Lambda}{\Lambda}
	\right)
	\right].
	\label{eq:app-mass-A2}
\end{equation}
Therefore, the first power-suppressed correction behaves as
$x^2A_2(x)\sim-x^2\Lambda^5/24$.

At order $x^4$, including all contributions and using the order-$x^2$
equation to simplify their sum gives
\begin{align}
	A_4(x)
	={}&
	\frac{1}{\displaystyle
		1+\frac{2\ell}{M_{\rm L}}}
	\Bigg\{
	\frac{1+\ell}{M_{\rm L}^2}A_2^2
	+\left[
	\frac12
	+M_{\rm L}
	\left(
	\frac1{24}
	-\frac{\ell}{12}
	-4C_2
	\right)
	\right]A_2
	\nonumber\\
	&\hspace{28mm}
	+\frac{M_{\rm L}}8
	+M_{\rm L}^4
	\left(
	\frac{\ell}{3840}
	-2C_4
	\right)
	\Bigg\}.
	\label{eq:app-mass-A4-exact}
\end{align}
Similarly using \eqref{eq:app-mass-leading-orders} and
\eqref{eq:app-mass-A2} in \eqref{eq:app-mass-A4-exact} shows that
$A_4=O(\Lambda^9)$.  It follows that the $x^4$ sector is smaller than
the $x^2$ sector by a factor of order $x^2\Lambda^4$, which vanishes as
$x\to0$.\\
%

Now every term carrying a positive power of $x$ lies beyond all fixed
orders in the inverse-logarithmic expansion of $M_{\rm L}(x)$.  For
this reason, the implicit form \eqref{eq:app-mass-leading-sector} is
preferable when evaluating the power-suppressed sectors.

Together, \eqref{eq:app-mass-leading-sector},
\eqref{eq:app-mass-A2-exact}, and \eqref{eq:app-mass-A4-exact}
determine the expansion \eqref{eq:app-mass-sector-ansatz}. Iterating
the leading-sector equation gives the following explicit expression
for $M_{\rm L}$, whose terms through order $\Lambda^0$ reproduce
\eqref{eq:mass-corrected} in the main text,
\begin{align}
	M_{\rm L}
	={}&
	\Lambda^2
	-2\Lambda\log\!\frac{\Lambda^2}{2}
	+\log^2\!\frac{\Lambda^2}{2}
	+4\log\!\frac{\Lambda^2}{2}
	+2C_0
	-
	\frac{
		2\log^2\!\frac{\Lambda^2}{2}
		+8\log\!\frac{\Lambda^2}{2}
		+4C_0
	}{\Lambda}
	+O\Big(
	\frac{\log^3\Lambda}{\Lambda^2}
	\Big).
	\label{eq:app-mass-ML-explicit}
\end{align}
\section{Non-uniformity of the $\zeta$-saddle in $O(N)$ model}
\label{Non-uniform}
In this appendix we demonstrate that for the conventional $O(N)$ model the leading large $N$ sector of partition function is not completely determined by the zero mode of the auxiliary field $ \zeta$ at non-zero angular potential. At zero angular potential the non-zero modes enters only at subleading orders at large $N$ as reviewed in the appendix C of \cite{David:2025tqn}. The presence of the non-zero angular potential breaks the rotational symmetry $O(3)$ to $O(2)$, leading to non-zero modes contribute at the leading order in large $N$. We show this explicitly in the following.

Let us consider the partition function for the $O(N)$ model as was given in \eqref{eq:HS-path-integral}
\begin{align}
	Z(\beta,\mu)
	={}&\int\mathcal D\zeta\,\mathcal D\phi\,
	\exp\Bigl\{-\frac12\int d^3x\,\sqrt g\,\Bigl[
	g^{\mu\nu}(D_\mu\phi_i)(D_\nu\phi_i)
	+\Big(\frac{\mathcal R}{8}-i\zeta\Big)\phi_i\phi_i
	+\frac{N}{4\lambda}\zeta^2
	\Bigr]\Bigr\}.
	\label{eq:HS-path-integral-appen}
\end{align}
Now one has to compute the saddle point of the path integral over the auxiliary field $\zeta$.  The saddle point condition in $ \zeta$ is given by 
\begin{align}\label{saddle cond}
	\frac{N\zeta}{2\lambda}=i\langle{\phi_i\phi_i}\rangle_{\beta}.
\end{align}
We separate the zero-mode and non-zero modes of the auxiliary field $\zeta$ as
\begin{align}
	\zeta(\Omega)=\zeta_{0}+\tilde \zeta(\Omega)
\end{align}
The $\zeta_0$ denotes the constant mode of $\zeta(\Omega)$. Integrating the both side of the saddle point condition \eqref{saddle cond} on $S^1\times S^2$ including the $\sqrt{g}$ factor, we have saddle point condition for the zero mode $\zeta_0$
\begin{align}\label{saddle cond zero}
	\frac{\zeta_0}{2\lambda}=\frac{i}{4\pi}\int d\Omega\,
	\langle \phi^2(\Omega)\rangle_{\beta}
\end{align}
where $d\Omega=\sin\theta d\theta d\varphi$.
Note the cancellation of the factor of $N$ from both sides of the equation as $\langle\phi_i\phi_i\rangle_{\beta}=N\langle\phi^2\rangle_{\beta}$, where $\phi$ denotes a single scalar component.
Now subtracting \eqref{saddle cond zero} from \eqref{saddle cond} we obtain the saddle point condition for the non-zero modes, as given below
\begin{align}\label{saddle cond non zero}
	\frac{\tilde\zeta(\Omega)}{2\lambda}
	=i\left[
	\langle \phi^2(\Omega)\rangle_{\beta}
	-\frac{1}{4\pi}\int d\Omega'\,
	\langle \phi^2(\Omega')\rangle_{\beta}
	\right].
\end{align}
Thus at the saddle point at large $N$ the auxiliary field has to satisfy the equations \eqref{saddle cond zero} and \eqref{saddle cond non zero} simultaneously.
We now ask whether this is possible within a restricted ansatz in which the auxiliary field contains only its zero mode $\zeta_0$, while all non-zero modes are set to zero. 
In this ansatz, the zero-mode at saddle point simply plays the role of an effective mass for the scalar fields. We can therefore evaluate $\langle\phi^2\rangle_{\beta}$ from the coincident limit of the corresponding massive scalar two-point function, after subtracting the ultraviolet-divergent vacuum contribution.
Then we can substitute this into the RHS of \eqref{saddle cond non zero} and  see that RHS computes to be non-zero. This is incompatible with the original assumption in the ansatz that all non-zero modes $\tilde\zeta$ vanish.\\

It therefore remains to evaluate $\langle \phi^2\rangle_{\beta}$ explicitly and verify that it indeed has a non-vanishing projection onto the non-zero-mode sector. We will carry out this calculation in the following, thereby establishing the inconsistency of the zero-mode-only ansatz.

%
%

\subsection*{Bare coincident two-point function}
With the ansatz for the auxiliary field $\zeta$ which only allows non-vanishing zero mode and all the non-zero modes are vanishing, we can easily find the two point function of the field $\phi$ as the non-vanishing zero mode acts only as mass of the scalar field $\phi$. More precisely $-i\zeta_0$ is identified as this mass and thus we have the notation
\begin{align}
	\tilde m^2=-i\zeta_0
\end{align}
For a single scalar component, the coincident two-point function in the constant-mass background is
\begin{equation}
	G_{\rm bare}(\Omega;\mu)
	=
	\frac{1}{\beta r^2}
	\sum_{n\in\mathbb Z}\sum_{\ell=0}^{\infty}\sum_{m=-\ell}^{\ell}
	\frac{|Y_{\ell m}(\Omega)|^2}
	{(\omega_n+\mu m)^2+E_\ell^2}.
	\label{eq:bare}
\end{equation}
where 
\[
E_\ell^2=\frac{(\ell+\tfrac12)^2}{r^2}+\tilde m^2,
\qquad
\omega_n=\frac{2\pi n}{\beta}.
\]
The Matsubara sum in \eqref{eq:bare} can be evaluated directly as
\[
\begin{aligned}
	\frac1\beta\sum_{n\in\mathbb Z}
	\frac1{(\omega_n+a)^2+E^2}
	&=
	\frac1{2iE\beta}\sum_{n\in\mathbb Z}
	\left[
	\frac1{\omega_n+a-iE}
	-
	\frac1{\omega_n+a+iE}
	\right]\\
	&=
	\frac1{4iE}
	\left[
	\cot\!\left(\frac{\beta(a-iE)}2\right)
	-
	\cot\!\left(\frac{\beta(a+iE)}2\right)
	\right]\\
	&=
	\frac1{2E}
	\frac{\sinh(\beta E)}{\cosh(\beta E)-\cos(\beta a)},
	\qquad E>0.
\end{aligned}
\]
Setting $a=\mu m$ and $E=E_\ell$ immediately gives
\begin{equation}
	G_{\rm bare}(\Omega;\mu)
	=
	\frac1{2r^2}
	\sum_{\ell,m}
	\frac{|Y_{\ell m}(\Omega)|^2}{E_\ell}
	\frac{\sinh(\beta E_\ell)}
	{\cosh(\beta E_\ell)-\cos(\beta\mu m)}.
	\label{eq:bare-summed}
\end{equation}
One can rewrite this as
\begin{equation}
	G_{\rm bare}(\Omega;\mu)
	=
	\frac1{2r^2}
	\sum_{\ell,m}
	\frac{|Y_{\ell m}(\Omega)|^2}{E_\ell}
	\Big(1+2\sum_{s=1}^{\infty}e^{-s\beta E_\ell}\cos(s\beta \mu m)\Big).
\end{equation}
Then \eqref{eq:bare-summed} splits as
\begin{equation}
	G_{\rm bare}(\Omega;\mu)=G_{\rm vac}^{\rm bare}(\tilde m)+G_{\rm th}(\Omega;\mu,\tilde m),
	\label{eq:split}
\end{equation}
where
\begin{align}
	G_{\rm vac}^{\rm bare}(\tilde m)
	&=
	\frac1{2r^2}\sum_{\ell,m}
	\frac{|Y_{\ell m}(\Omega)|^2}{E_\ell}
	=
	\frac1{8\pi r^2}\sum_{\ell=0}^{\infty}\frac{2\ell+1}{E_\ell},
	\label{eq:vac}\\[2mm]
	G_{\rm th}(\Omega;\mu,\tilde m)
	&=
	\frac1{r^2}
	\sum_{\ell=0}^{\infty}\sum_{m=-\ell}^{\ell}
	\frac{|Y_{\ell m}(\Omega)|^2}{E_\ell}
	\sum_{s=1}^{\infty}e^{-s\beta E_\ell}\cos(s\beta\mu m).
	\label{eq:thermal}
\end{align}
The vacuum term \eqref{eq:vac} contains the ultraviolet divergence, but is independent of
$\Omega$ and of $\mu$.  The  term \eqref{eq:thermal} is finite.

The finite thermal expectation value is obtained by subtracting the vacuum
short-distance singularity from the two-point function before taking the
coincident limit.  Denoting this prescription by $|_{\rm ren}$ and the finite
vacuum remainder by $G_{\rm vac}^{\rm ren}(\tilde m)$, we write
\begin{equation}
	\langle\phi^2(\Omega)\rangle_{\beta}
	=\left.G_{\rm bare}(\Omega;\mu)\right|_{\rm ren}
	=G_{\rm vac}^{\rm ren}(\tilde m)+G_{\rm th}(\Omega;\mu,\tilde m).
	\label{eq:ren-def}
\end{equation}
The subtraction is already implicit in $\langle\phi^2\rangle_{\beta}$, whose
subscript labels the thermal state.  This expectation value includes both
the finite vacuum remainder and the thermal contribution.
We will not need the explicit value of $G_{\rm vac}^{\rm ren}$; the uniform gap equation discussed below will determine it relative to the thermal contribution.

\subsection*{Uniform gap equation}

For a constant mass, varying the renormalized scalar determinant with respect to
$\tilde m^2$ gives
\begin{equation}
	\frac{\partial\log Z}{\partial\tilde m^2}
	=-\frac12\int_0^\beta\dd\tau\int_{S^2}r^2\dd\Omega\,
	\langle\phi^2(\Omega)\rangle_{\beta}.
	\label{eq:deriv}
\end{equation}
At the critical large-$N$ homogeneous saddle the left-hand side vanishes as also can be seen from \eqref{saddle cond zero} at $\lambda\to \infty$.  Therefore
\begin{equation}
	\int_{S^2}\dd\Omega\,\langle\phi^2(\Omega)\rangle_{\beta}=0.
	\label{eq:integrated-gap}
\end{equation}
It is convenient to define the angular average
\begin{equation}
	\overline{G}_{\rm th}(\mu,\tilde m)
	\equiv
	\frac1{4\pi}\int_{S^2}\dd\Omega\,G_{\rm th}(\Omega;\mu,\tilde m).
	\label{eq:average}
\end{equation}
Using \eqref{eq:ren-def} in \eqref{eq:integrated-gap} gives, at the homogeneous saddle,
\begin{equation}
	G_{\rm vac}^{\rm ren}(\tilde m_{})
	+\overline{G}_{\rm th}(\mu,\tilde m)=0.
	\label{eq:gap-average}
\end{equation}
Substituting this back into \eqref{eq:ren-def} gives the exact and very useful relation
\begin{equation}
	\left.\langle\phi^2(\Omega)\rangle_{\beta}\right|_{\rm hom}
	=
	G_{\rm th}(\Omega;\mu,\tilde m)-\overline{G}_{\rm th}(\mu,\tilde m).
	\label{eq:key}
\end{equation}
No expansion in $\mu$ has been made in obtaining \eqref{eq:key}.

\subsection*{Zero angular potential: $\mu=0$}

When $\mu=0$, the factor $\cos(s\beta\mu m)$ in \eqref{eq:thermal} is equal to one and therefore does not depend on $m$.  Hence
\begin{align}
	G_{\rm th}(\Omega;0,\tilde m)
	&=
	\frac1{r^2}
	\sum_{\ell=0}^{\infty}\frac1{E_\ell}
	\sum_{s=1}^{\infty}e^{-s\beta E_\ell}
	\sum_{m=-\ell}^{\ell}|Y_{\ell m}(\Omega)|^2
	\notag\\
	&=
	\frac1{4\pi r^2}
	\sum_{\ell=0}^{\infty}
	\frac{2\ell+1}{E_\ell}
	\sum_{s=1}^{\infty}e^{-s\beta E_\ell}.
	\label{eq:mu0-thermal}
\end{align}
The right-hand side is independent of $\Omega$.  Therefore
\begin{equation}
	G_{\rm th}(\Omega;0,\tilde m)=\overline{G}_{\rm th}(0,\tilde m).
	\label{eq:mu0-constant}
\end{equation}
Equation \eqref{eq:key} then immediately gives
\begin{equation}
	\left.\langle\phi^2(\Omega)\rangle_{\beta}\right|_{\mu=0,\,\rm hom}=0
	\qquad\text{for every }\Omega.
	\label{eq:mu0-zero}
\end{equation}
Thus at $\mu=0$ the integrated uniform gap equation is also the local gap equation, because rotational symmetry makes the expectation value constant over the sphere.

\subsection*{Nonzero angular potential: small-$\mu$ comparison}

For $\mu\neq0$, the factor $\cos(s\beta\mu m)$ in \eqref{eq:thermal} depends on $m$, so the thermal expectation value need not be independent of position.  To display this in the simplest way, expand only this factor for small Euclidean angular potential:
\begin{equation}
	\cos(s\beta\mu m)
	=1-\frac12 s^2\beta^2\mu^2m^2+O(\mu^4).
	\label{eq:cos-smallmu}
\end{equation}
Substituting \eqref{eq:cos-smallmu} into \eqref{eq:thermal} gives
\begin{align}
	G_{\rm th}(\Omega;\mu,\tilde m)
	={}&G_{\rm th}(\Omega;0,\tilde m)
	-\frac{\beta^2\mu^2}{2r^2}
	\sum_{\ell=0}^{\infty}\frac1{E_\ell}
	\sum_{s=1}^{\infty}s^2e^{-s\beta E_\ell}
	\sum_{m=-\ell}^{\ell}m^2|Y_{\ell m}(\Omega)|^2
	+O(\mu^4).
	\label{eq:thermal-smallmu-1}
\end{align}
The first term is the position-independent quantity already obtained at $\mu=0$.  For the second term we use
\begin{equation}
	\sum_{m=-\ell}^{\ell}m^2|Y_{\ell m}(\theta,\varphi)|^2
	=
	\frac{(2\ell+1)\ell(\ell+1)}{8\pi}\sin^2\theta.
	\label{eq:m2identity}
\end{equation}
Therefore
\begin{equation}
	G_{\rm th}(\theta;\mu,\tilde m)
	=G_{\rm th}(0,\tilde m)-\mu^2\mathcal A(\beta,r,\tilde m)\sin^2\theta+O(\mu^4).
	\label{eq:gth-theta}
\end{equation}
Where $\mathcal A(\beta,r,\tilde m)$ is given by
on $\tilde m$ and is given by
\begin{equation}
	\mathcal A(\beta,r,\tilde m)
	=
	\frac{\beta^2}{16\pi r^2}
	\sum_{\ell=1}^{\infty}
	\frac{(2\ell+1)\ell(\ell+1)}{E_\ell}
	\sum_{s=1}^{\infty}s^2e^{-s\beta E_\ell}>0.
	\label{eq:Adef}
\end{equation}
The angular average of \eqref{eq:gth-theta} can be done using
\begin{equation}
	\frac1{4\pi}\int_{S^2}\dd\Omega\,\sin^2\theta=\frac23,
	\label{eq:sinaverage}
\end{equation}
and hence
\begin{equation}
	\overline G_{\rm th}(\mu,\tilde m)
	=G_{\rm th}(0,\tilde m)-\frac23\mu^2\mathcal A(\beta,r,\tilde m)+O(\mu^4).
	\label{eq:gth-average-smallmu}
\end{equation}
Now use the homogeneous gap equation through the exact relation \eqref{eq:key}.  The finite local thermal expectation value at the homogeneous saddle is
\begin{align}
	\left.\langle\phi^2(\theta)\rangle_{\beta}\right|_{\rm hom}
	&=G_{\rm th}(\theta;\mu,\tilde m)-\overline G_{\rm th}(\mu,\tilde m)
	\notag\\
	&=-\mu^2\mathcal A(\beta,r,\tilde m)
	\left(\sin^2\theta-\frac23\right)+O(\mu^4).
	\label{eq:phi2-smallmu}
\end{align}
This immediately shows both facts
\begin{equation}
	\frac1{4\pi}\int\dd\Omega\,
	\left.\langle\phi^2(\Omega)\rangle_{\beta}\right|_{\rm hom}=0,
\end{equation}
but, for generic $\theta$ and nonzero $\mu$,
\begin{equation}
	\left.\langle\phi^2(\theta)\rangle_{\beta}\right|_{\rm hom}\neq0.
\end{equation}
Thus the homogeneous gap equation still kills the angular average, but it does not kill the local expectation value once the angular potential breaks the full rotational symmetry.

To make this result explicit in the small-$\beta/r$ regime, we keep
$\tilde m\beta>0$ fixed.
Applying the midpoint Euler--Maclaurin expansion to \eqref{eq:Adef} gives
\begin{equation}
	\mathcal A(\beta,r,\tilde m)
	=
	\frac{r^2}{4\pi\beta}
	\left[
	\frac{\tilde m\beta}{\e^{\tilde m\beta}-1}
	-\log\!\left(1-\e^{-\tilde m\beta}\right)
	-\frac{1}{32\sinh^2(\tilde m\beta/2)}\left(\frac{\beta}{r}\right)^2
	+O\!\left(\frac{\beta^4}{r^4}\right)
	\right].
	\label{eq:A-small-beta}
\end{equation}
Substituting \eqref{eq:A-small-beta} into \eqref{eq:phi2-smallmu} gives
\begin{align}
	\left.\langle\phi^2(\theta)\rangle_{\beta}\right|_{\rm hom}
	={}&
	-\frac{\mu^2r^2}{4\pi\beta}
	\left[
	\frac{\tilde m\beta}{\e^{\tilde m\beta}-1}
	-\log\!\left(1-\e^{-\tilde m\beta}\right)
	-\frac{1}{32\sinh^2(\tilde m\beta/2)}\left(\frac{\beta}{r}\right)^2
	+O\!\left(\frac{\beta^4}{r^4}\right)
	\right]\notag\\
	&\quad\times\left(\sin^2\theta-\frac23\right)
	+O(\mu^4).
	\label{eq:phi2-small-beta}
\end{align}

On the homogeneous ansatz $\tilde\zeta=0$.  Using
\eqref{eq:phi2-smallmu}, whose angular average vanishes, the right-hand side
of \eqref{saddle cond non zero} becomes
\begin{equation}
	\left.
	i\left[
	\langle\phi^2(\theta)\rangle_{\beta}
	-\frac{1}{4\pi}\int d\Omega\,
	\langle\phi^2(\Omega)\rangle_{\beta}
	\right]\right|_{\rm hom}
	=
	-i\mu^2\mathcal A(\beta,r,\tilde m)
	\left(\sin^2\theta-\frac23\right)
	+O(\mu^4).
\end{equation}
For $\mu\neq0$, this expression is nonzero for generic $\theta$ because
$\mathcal A(\beta,r,\tilde m)>0$.  However, the left-hand side of
\eqref{saddle cond non zero} vanishes when $\tilde\zeta=0$.  Therefore the
homogeneous ansatz does not satisfy the full saddle equation, and nonzero
modes of $\zeta$ must be included at leading order.
\section*{ The pp-wave geometry}

We now apply the same test to the usual $O(N)$ model on the pp-wave
geometry \eqref{ppwave}.  We first separate the auxiliary-field  into
its constant and non-zero projections.  We then set all nonconstant modes
to zero, compute the coincident scalar propagator in the resulting
constant-mass background, and substitute it into the nonconstant equation.
The purpose is to determine whether this ansatz satisfies the full saddle
equation.

After continuation to Euclidean time, the metric and volume element are
\begin{equation}
	\dd s_E^2=(1+y^2)\dd\tau^2+r^2\dd y^2-i r\,\dd x\,\dd\tau,
	\qquad \tau\sim\tau+\beta,
	\qquad \sqrt g=\frac{r^2}{2}.
	\label{eq:ppmetric-appendix}
\end{equation}
The Hubbard--Stratonovich action is \eqref{eq:HS-path-integral-appen} with
this metric and $\mathcal R=0$.  Its local saddle equation is given by
\begin{equation}
	\frac{\zeta(\tau,x,y)}{2\lambda}
	=i\langle\phi^2(\tau,x,y)\rangle_{\beta}.
	\label{eq:pplocalsaddle}
\end{equation}
Here $\phi$ denotes one scalar component, so the common factor of $N$ has
cancelled from the two sides.

We restrict to configurations that are independent of $\tau$ and $x$, so
that $\zeta=\zeta(y)$.  Because the $y$ direction is noncompact, we define
the constant and nonconstant projections using an average over a finite
interval $[-Y,Y]$
\begin{equation}
	\overline F^{\,(Y)}\equiv\frac1{2Y}\int_{-Y}^{Y}\dd y\,F(y),
	\qquad
	\zeta(y)=\zeta_0+\tilde\zeta_{0,0}(y),
	\qquad
	\overline{\tilde\zeta}_{0,0}^{\,(Y)}=0.
	\label{eq:ppprojection}
\end{equation}
The interval is used only to define the projection; the propagator is still
evaluated on the full $y$ line.  Averaging \eqref{eq:pplocalsaddle} over this
interval gives the constant equation, and subtracting the average from the
local equation gives the nonconstant equation
\begin{align}
	\frac{\zeta_0}{2\lambda}
	&=i\,\overline{\langle\phi^2\rangle_{\beta}}^{\,(Y)},
	\label{eq:ppzerosaddle}\\
	\frac{\tilde\zeta_{0,0}(y)}{2\lambda}
	&=i\left[
	\langle\phi^2(y)\rangle_{\beta}
	-\overline{\langle\phi^2\rangle_{\beta}}^{\,(Y)}
	\right].
	\label{eq:ppnonzerosaddle}
\end{align}
They are the pp-wave counterparts of \eqref{saddle cond zero} and
\eqref{saddle cond non zero}, and must hold simultaneously.  At criticality,
$\lambda\to\infty$, the first equation requires the averaged expectation
value to vanish, while the second requires its nonconstant part to vanish.

We now impose the ansatz
\begin{equation}
	\zeta(\tau,x,y)=\zeta_0,
	\qquad \tilde\zeta(\tau,x,y)=0,
	\qquad \hat m^2=-i\zeta_0,
	\qquad \hat m\geq0.
	\label{eq:ppuniformansatz}
\end{equation}
The question is whether
the right-hand side of \eqref{eq:ppnonzerosaddle} vanishes in this
constant-mass background, as its left-hand side does.  We do not need to
assume that the averaged equation has a solution in order to perform this
test.

\subsubsection*{Bare coincident two-point function}

As earlier, we take $k\geq0$ and include the complex-conjugate
modes.  Let $\omega_n=2\pi n/\beta$.  For $k>0$, the normalized transverse
oscillator modes are $\psi_{\ell,k}(y)$, with energies
\begin{equation}
	E_{\ell k}\equiv\ell+\frac12+k+\frac{\hat m^2r^2}{4k}.
	\label{eq:ppE}
\end{equation}
At $k=0$, the oscillator potential vanishes and the $y$ dependence is
described by the free-particle modes
\begin{equation}
	\psi_{q,0}(y)=\frac{e^{iqy}}{\sqrt{2\pi}},
	\qquad q\in\mathbb R.
\end{equation}
Thus, keeping a time separation $u=\tau-\tau'$, the two-point function can
be written directly on the noncompact $x$ line as
\begin{align}
	G_{\rm bare}^{\rm pp}(u,y;\hat m)
	&=\frac2\beta\sum_{n\in\mathbb Z}e^{i\omega_n u}
	\int_0^\infty\frac{\dd k}{2\pi}\,\mathcal K_{n,k}(y),
	\notag\\
	\mathcal K_{n,k}(y)
	&=
	\begin{cases}
		\displaystyle
		\sum_{\ell=0}^{\infty}\frac{|\psi_{\ell,k}(y)|^2}{4k}
		\left[
		\dfrac1{E_{\ell k}+i r\omega_n}
		+\dfrac1{E_{\ell k}-i r\omega_n}
		\right],&k>0,\\[4mm]
		\displaystyle
		\int_{\mathbb R}\frac{\dd q}{2\pi}
		\frac1{q^2+\hat m^2r^2},&k=0.
	\end{cases}
	\label{eq:ppbare}
\end{align}
The second line keeps the $k=0$ free-particle sector explicit.  Define its $n$-independent spatial kernel
by
\begin{equation}
	\mathcal I_0(\hat m)
	\equiv\mathcal K_{n,0}(y)
	=\int_{\mathbb R}\frac{\dd q}{2\pi}
	\frac1{q^2+\hat m^2r^2}.
	\label{eq:ppkzero-kernel}
\end{equation}
At finite temperature, the corresponding frequency sum is
\begin{align}
	\mathcal C_0^{(\beta)}(u;\hat m)
	&\equiv
	\frac1\beta\sum_{n\in\mathbb Z}e^{i\omega_nu}\,
	\mathcal I_0(\hat m)
	=\delta_\beta(u)\mathcal I_0(\hat m),
	\notag\\
	\delta_\beta(u)
	&\equiv\sum_{p\in\mathbb Z}\delta(u-p\beta).
	\label{eq:ppkzero-finiteT}
\end{align}
The zero-temperature counterpart to the above frequency integral gives
\begin{equation}
	\mathcal C_0^{(0)}(u;\hat m)
	\equiv
	\int_{-\infty}^{\infty}\frac{\dd\omega}{2\pi}
	e^{i\omega u}\mathcal I_0(\hat m)
	=\delta(u)\mathcal I_0(\hat m).
	\label{eq:ppkzero-zeroT}
\end{equation}
Therefore the vacuum-subtracted $k=0$ contribution is
\begin{align}
	\mathcal C_0^{\rm th}(u;\hat m)
	&\equiv
	\mathcal C_0^{(\beta)}(u;\hat m)
	-\mathcal C_0^{(0)}(u;\hat m)
	\notag\\
	&=\mathcal I_0(\hat m)
	\sum_{p\in\mathbb Z\setminus\{0\}}\delta(u-p\beta)
	=0,
	\qquad |u|<\beta.
	\label{eq:ppkzero-subtracted}
\end{align}
In particular, its coincidence limit $u\to0$ vanishes after the vacuum
subtraction.  This is why no separate $k=0$ term appears further in the analysis.  The continuum limit $k\to0^+$ is not removed and
remains included in every subsequent $k$-integral.

The prefactor in \eqref{eq:ppbare} includes $\sqrt g=r^2/2$; all sums and
integrals are understood with regulators until the vacuum subtraction is
performed.  For $k>0$, the paired Matsubara sum at coincidence gives
\begin{equation}
	\frac1\beta\sum_{n\in\mathbb Z}
	\left[\frac1{E+i r\omega_n}+\frac1{E-i r\omega_n}\right]
	=\frac1r\coth\!\left(\frac{\beta E}{2r}\right),
	\qquad E>0.
	\label{eq:ppmatsu}
\end{equation}
After separating the $k=0$ contact term described above, the non-contact
part of the coincident propagator is
\begin{equation}
	G_{\rm bare}^{\rm pp}(y;\hat m)
	=\frac1{2r}\int_0^\infty\frac{\dd k}{2\pi k}
	\sum_{\ell=0}^{\infty}|\psi_{\ell,k}(y)|^2
	\coth\!\left(\frac{\beta E_{\ell k}}{2r}\right).
	\label{eq:ppbaresummed}
\end{equation}
Using $\coth(z/2)=1+2/(e^z-1)$ separates the vacuum and thermal terms,
\begin{equation}
	G_{\rm bare}^{\rm pp}(y;\hat m)
	=G_{\rm vac}^{\rm bare}(\hat m)
	+G_{\rm th}^{\rm pp}(y;\hat m),
	\label{eq:ppsplit}
\end{equation}
where the two terms are explicitly
\begin{equation}
	\begin{aligned}
		G_{\rm vac}^{\rm bare}(\hat m)
		&=\frac1{2r}\int_0^\infty\frac{\dd k}{2\pi k}
		\sum_{\ell=0}^{\infty}|\psi_{\ell,k}(y)|^2,\\
		G_{\rm th}^{\rm pp}(y;\hat m)
		&=\frac1r\int_0^\infty\frac{\dd k}{2\pi k}
		\sum_{\ell=0}^{\infty}
		\frac{|\psi_{\ell,k}(y)|^2}{e^{\beta E_{\ell k}/r}-1}.
	\end{aligned}
	\label{eq:ppthermalmode}
\end{equation}
The vacuum term is ultraviolet divergent and must be renormalized.  For a
constant mass, its renormalized finite part is independent of $y$, whereas
the thermal term is finite and contains all the $y$ dependence.  Therefore
\begin{equation}
	\langle\phi^2(y)\rangle_{\beta}
	=\left.G_{\rm bare}^{\rm pp}(y;\hat m)\right|_{\rm ren}
	=G_{\rm vac}^{\rm ren}(\hat m)
	+G_{\rm th}^{\rm pp}(y;\hat m).
	\label{eq:ppren}
\end{equation}
As on the sphere, the explicit finite vacuum remainder is unnecessary for
the nonconstant projection, since it cancels from that equation.

\subsubsection*{Evaluation of the thermal contribution}

Expand the Bose factor in \eqref{eq:ppthermalmode} as a series in the corresponding exponential and
write $a_s=s\beta/r$.  The diagonal Mehler formula gives
\begin{equation}
	\sum_{\ell=0}^{\infty}e^{-a_s(\ell+1/2)}|\psi_{\ell,k}(y)|^2
	=\sqrt{\frac{k}{\pi\sinh a_s}}
	e^{-2ky^2\tanh(a_s/2)}.
	\label{eq:ppmehler}
\end{equation}
Define, for use in the following expressions,
\begin{equation}
	a_s\equiv\frac{s\beta}{r},
	\qquad C_s(y)\equiv a_s+2y^2\tanh\!\left(\frac{a_s}{2}\right)>0.
	\label{eq:ppthermalparameters}
\end{equation}
Then the thermal term becomes
\begin{equation}
	G_{\rm th}^{\rm pp}(y;\hat m)
	=\frac1{2\pi r}\sum_{s=1}^{\infty}\frac1{\sqrt{\pi\sinh a_s}}
	\int_0^\infty\dd k\,k^{-1/2}
	\exp\!\left[-C_s(y)k-\frac{a_s\hat m^2r^2}{4k}\right].
	\label{eq:ppbeforek}
\end{equation}
Performing the integral over $k$ in the above equation we have
\begin{equation}
	G_{\rm th}^{\rm pp}(y;\hat m)
	=\frac1{2\pi r}\sum_{s=1}^{\infty}
	\frac{\exp\!\left[-r\hat m\sqrt{a_s C_s(y)}\right]}
	{\sqrt{\sinh a_s}\sqrt{C_s(y)}}.
	\label{eq:ppthermalexplicit}
\end{equation}
Every term is positive and strictly decreases as $y^2$ increases.  This
holds also at $\hat m=0$, when the exponential factor in each summand equals
one.  In particular,
\begin{equation}
	G_{\rm th}^{\rm pp}(0;\hat m)
	>G_{\rm th}^{\rm pp}(y;\hat m),
	\qquad y\neq0,\quad\hat m\geq0.
	\label{eq:ppnonuniform}
\end{equation}
The thermal coincident function is therefore nonconstant at every finite
$\beta/r>0$ in the range of constant masses under consideration.

The small-$\beta/r$ behavior can also be obtained directly from this mode
sum.  For fixed $y$ and fixed $\hat m\beta>0$, expanding
$\sinh a_s$ and $C_s(y)$ gives
\begin{equation}
	G_{\rm th}^{\rm pp}(y;\hat m)
	=-\frac{\log\!\left(1-e^{-\hat m\beta\sqrt{1+y^2}}\right)}
	{2\pi\beta\sqrt{1+y^2}}
	+O\!\left(\frac{\beta}{r^2}\right).
	\label{eq:ppthermalhighT}
\end{equation}
Thus the position dependence is present already at leading order in this
expansion.  The massless case must instead be taken in the exact expression
\eqref{eq:ppthermalexplicit}, since the fixed-$\hat m\beta>0$ expansion
is not uniform as $\hat m\beta\to0$.

\subsubsection*{Testing the nonconstant saddle equation}

Define the averaged thermal contribution by
\begin{equation}
	\overline G_{\rm th}^{\,(Y)}(\hat m)
	\equiv\frac1{2Y}\int_{-Y}^{Y}\dd y\,
	G_{\rm th}^{\rm pp}(y;\hat m).
	\label{eq:ppavg}
\end{equation}
If a constant background satisfies the averaged critical equation
\eqref{eq:ppzerosaddle}, it obeys
\begin{equation}
	G_{\rm vac}^{\rm ren}(\hat m)
	+\overline G_{\rm th}^{\,(Y)}(\hat m)=0.
	\label{eq:ppgapavg}
\end{equation}
Substituting this into \eqref{eq:ppren} gives precisely the same logical
relation as \eqref{eq:key}
\begin{equation}
	\left.\langle\phi^2(y)\rangle_{\beta}\right|_{\rm hom}
	=G_{\rm th}^{\rm pp}(y;\hat m)
	-\overline G_{\rm th}^{\,(Y)}(\hat m).
	\label{eq:ppkey}
\end{equation}
The right-hand side has vanishing average, but it is not identically zero.
More directly, whether or not the averaged equation has been imposed, the
nonconstant saddle residual on the ansatz \eqref{eq:ppuniformansatz} is
\begin{equation}
	\left.i\left[
	\langle\phi^2(y)\rangle_{\beta}
	-\overline{\langle\phi^2\rangle_{\beta}}^{\,(Y)}
	\right]\right|_{\tilde\zeta=0}
	=i\left[G_{\rm th}^{\rm pp}(y;\hat m)
	-\overline G_{\rm th}^{\,(Y)}(\hat m)\right].
	\label{eq:ppnonzeroresidual}
\end{equation}
At the center of any interval with $Y>0$, \eqref{eq:ppnonuniform} implies
\begin{equation}
	G_{\rm th}^{\rm pp}(0;\hat m)
	-\overline G_{\rm th}^{\,(Y)}(\hat m)>0.
	\label{eq:ppcenterpositive}
\end{equation}
Hence the right-hand side of \eqref{eq:ppnonzerosaddle} is nonzero, whereas
its left-hand side vanishes on the assumed configuration.  Equivalently,
one may choose a smooth transverse variation with zero integral, supported
near two points where \eqref{eq:ppthermalexplicit} differs; its contraction
with the local expectation value is nonzero.  

We have therefore shown that a configuration containing only $\zeta_0$
does not satisfy the full local saddle equation of the conventional $O(N)$
model on the thermal pp-wave.  The uncancelled source lies in the static,
$x$-independent sector $\tilde\zeta_{0,0}(y)$, which is a nonconstant mode
despite having $k=0$.  One can also show that its contribution to the first variation of the
effective action is proportional to $N$, so it cannot be discarded as a
subleading fluctuation.  This establishes the inconsistency of the uniform
ansatz for the usual $O(N)$ model both on pp wave geometry and on $S^1\times S^2$; it does not determine the
nonuniform saddle or its free energy.

\bibliographystyle{JHEP}
\bibliography{references}

\end{document}